\documentstyle [fleqn,12pt,epsf] {article} 
\newcommand{\al}{\alpha}

\newcommand{\de}{\delta}

\newcommand{\ep}{\epsilon}
\newcommand{\ga}{\gamma}
\newcommand{\Ga}{\Gamma}

\newcommand{\Om}{\Omega}

\newcommand{\be}{\begin{equation}}
\newcommand{\ee}{\end{equation}}
\newcommand{\bea}{\begin{eqnarray}}
\newcommand{\eea}{\end{eqnarray}}
\newcommand{\bean}{\begin{eqnarray*}}
\newcommand{\eean}{\end{eqnarray*}}
\newcommand{\dd}{\partial}

\newcommand{\gsim}{\stackrel{>}{\sim}}
\newcommand{\lsim}{\stackrel{<}{\sim}}

\begin{document}
\centerline{\large\bf Microwave Background Anisotropies}
\centerline{\large\bf from Scaling Seed 
	Perturbations}
\vskip24pt
\baselineskip=10pt
\centerline{Ruth Durrer and Mairi Sakellariadou}

\vskip10pt
\centerline{\it D\'epartement de Physique Th\'eorique}
\centerline{\it Universit\'e de Gen\`eve}
\centerline{\it 24 quai Ernest Ansermet, CH-1211 Gen\`eve 4, Switzerland}

\vskip 44pt
\begin{abstract}
\noindent
We study microwave background anisotropies induced by scaling seed 
perturbations in a universe dominated by cold dark matter. 
Using a gauge invariant linear perturbation analysis, we solve the perturbation
equations on super-horizon scales, for  CMB anisotropies
 triggered by generic gravitational seeds. 
We find that perturbations induced by seeds --- under very mild restrictions 
--- are nearly isocurvature. Thus, compensation, which is mainly the
consequence of physically sensible initial conditions, is very generic.

We then restrict our study to the case of 
scaling sources, motivated by global scalar fields. We parameterize the
energy momentum tensor of the source by ``seed functions'' and
calculate  the  Sachs-Wolfe and acoustic contributions to the CMB
anisotropies.
We  discuss the dependence of the anisotropy spectrum on the parameters of the 
 model considered.  Even within the restricted class of models
investigated in this work, we find a surprising variety of results for
the position and height of the first acoustic  peak as well as for the
overall amplitude. In particular, for certain choices of parameters,
the spectrum resembles very much the well known adiabatic inflationary
spectrum, whereas for others, the position of the first acoustic peak
is significantly shifted towards smaller angular scales.

\vskip 10pt
\vspace{1cm}
\end{abstract}
\vspace{1cm}
\vspace{1cm}

\section{Introduction}
The origin of the large scale structure in the universe, is clearly 
one of the most important open questions in cosmology.  Within the framework
of gravitational instability, there are two currently investigated
families of models to 
explain the formation of the observed structure.  Initial density 
perturbations can either be due to ``freezing in'' of quantum fluctuations
of a scalar field during an inflationary period 
\cite{stein}, or they may be seeded by topological defects, which can form 
naturally during a symmetry breaking phase transition in the early universe
\cite{kibble}. Inflationary fluctuations are produced at a very 
early stage of the evolution of the universe, and are driven far beyond the 
Hubble radius by inflationary expansion. Subsequently, they are not
altered anymore and evolve freely according to homogeneous linear
perturbation equations until late times. These fluctuations are termed
``passive'' and ``coherent'' \cite{joao}. ``Passive'', since no new
perturbations are created after inflation; ``coherent'' since randomness
only enters the creation of perturbations during inflation,
subsequently they evolve in a deterministic and coherent manner.

On the other hand, in models with topological defects or other types
of seeds, fluctuations are  generated continuously and evolve
according to inhomogeneous linear perturbation equations. The seeds are
 any   non--uniformly distributed form of energy, which 
contributes only a small fraction to the total energy density of the universe
and which interacts with the cosmic fluid only gravitationally.
We will be particularly interested in the case of global topological defects, 
playing the role of seeds.  
The energy momentum tensor of the seed  is determined by the defect 
 (seed) evolution which, in general,
 is a non-linear process. These perturbations are called ``active'' 
and ``incoherent'' \cite{joao}. ``Active'' since new fluid perturbations are
induced continuously due to the presence of the seeds;
 ``incoherent'' since the randomness of the
non-linear seed evolution which sources the perturbations can destroy
the coherence of  fluctuations in the cosmic fluid.

The cosmic microwave background (CMB) anisotropies provide a link 
between theoretical predictions and observational data, which may allow us
to distinguish between inflationary models and defect scenarios, by purely
linear analysis.   On large angular scales, both families of models 
predict an approximately scale-invariant Harrison-Zel'dovich spectrum
\cite{h,z}. Although, perturbations from defect models are 
 non-Gaussian, this signature is probably rather weak, especially on
 large scales, where cosmic variance is substantial, and its
observation might be quite difficult.

Acoustic  peaks have been  
extensively studied in inflationary models, where observations of  amplitude
and position of the peaks can be used to determine cosmological
parameters \cite{husu}. Some studies of simplified models where
perturbations are seeded by topological defects, have already appeared
in the literature\cite{ram,rem}.

In this paper, we present a general investigation of acoustic peaks 
for models with
active perturbations. We estimate the Sachs-Wolfe and acoustic 
contributions. Our ``seed functions'', which determine the energy
momentum tensor of the source are  motivated from 3d numerical
simulations of $\pi_3$ defects, textures \cite{turok,ruthzhou}, in a universe 
dominated by cold dark matter (CDM).  We restrict ourselves to scalar 
perturbations.  

In section 2, 
we study CMB anisotropies triggered by generic gravitational seeds.  In 
particular, we present the equation for the coefficients $C_\ell$ and 
discuss the Sachs-Wolfe contribution, the acoustic peaks and Silk
damping.  We  solve the perturbation equations
for super-horizon scales. Studying the Bardeen potentials $\Psi$
and $\Phi$, which describe the scalar geometry perturbations, we find that 
compensation is automatically obtained, i.e., that perturbations are nearly 
isocurvature.  
In section 3, we restrict our investigation 
to the case of global scalar fields, for which we deduce the power spectra
of the seed functions from numerical simulations and analytical scaling
 arguments. The results obtained there apply for general scaling sources.
In section 4, we present some numerical examples and discuss how the 
characteristics of the acoustic peaks depend on the model. We
summarize our conclusions in section 5.

{\bf Notation:} The Friedmann metric is given by
$a^2(-dt^2+\ga_{ij}dx^idx^j)$, where $a$  denotes the scale factor, $t$
is conformal time and $\gamma$ is the metric of a three space with
constant curvature $K$. We shall consider a universe dominated by cold 
dark matter and discuss the case $K=0$ exclusively.
 An over-dot stands for derivative with
respect to conformal time $t$, while prime denotes the derivative with
respect to $kt\equiv x$. Subscripts (or superscripts) $_r$ ($^{(r)}$) 
and $_c$ ($^{(c)}$)  indicate the 
baryon-radiation plasma and CDM, respectively.

\vspace{0.5cm}
\section{CMB anisotropies from seeds}
\vspace{0.5cm}
 The coefficients $C_\ell$ represent the angular power spectrum of 
 CMB anisotropies. They can be given in terms of  the
expansion of the angular correlation function
\[ \langle{\delta T\over T}({\bf n}){\delta T\over T}({\bf n}')
\rangle\left|_{{~}_{\!\!({\bf n\cdot n}'=\cos\vartheta)}}\right. =
  {1\over 4\pi}\sum_\ell(2\ell+1)C_\ell P_\ell(\cos\vartheta)~. \]
We want to investigate these coefficients in models where the fluctuations 
 are induced by seeds.
We restrict ourselves to scalar perturbations, but this  analysis is easily
extended  to include vector and tensor contributions. 
	
If we neglect  Silk damping in a first step and integrate the photon
geodesics in the perturbed metric,
gauge invariant linear perturbation analysis leads to \cite{d90,RuthReview}   
\begin{eqnarray} 
{\delta T\over T}({\bf x}, {\bf n}) &=& \left[
 - {1\over 4}D_g^{(r)}({\bf x}) -V_j({\bf x})n^j 
	-(\Psi-\Phi)({\bf x})\right]_i^f  \nonumber\\  &&
	+ \int_i^f (\dot{\Psi} - \dot{\Phi} )({\bf x}',t') d\tau  ~, 
\label{dT} 
\end{eqnarray}
where over-dot denotes derivative with respect to conformal time $t$.
$\Phi$ and $\Psi$ are the Bardeen potentials, quantities describing the 
perturbations in the geometry, $\bf V$ is the peculiar velocity of 
the baryon fluid with respect to the overall Friedmann expansion and
$D_g^{(r)}$ specifies the intrinsic density fluctuation in the radiation 
fluid.  There are several gauge invariant variables which describe density 
fluctuations; they all differ substantially on super-horizon scales
but  coincide  inside the horizon. $D_g$ corresponds to the density
fluctuation in the so-called ``flat slicing'', where the perturbation
of the 3-dimensional Riemann scalar vanishes.  The initial time is the time of
decoupling, $t_{dec}$ of baryons and radiation, which occured at a 
redshift of $z_{dec}\sim 1100$.

The final time values in the square bracket of Eq. (\ref{dT}) give
rise only to monopole contributions and the dipole due to our motion
with respect to the CMB and are disregarded in the following. 
Taking the Fourier transform of Eq. (\ref{dT}), we then obtain  
\begin{eqnarray}
{\delta T\over T} ({\bf k}, {\bf n}) = e^{i({\bf k} \cdot {\bf n})t_0} [
{1\over 4}D_g^{(r)}({\bf k}, t_{dec}) + {\bf V}({\bf k}, t_{dec})\cdot {\bf n}
+ (\Psi -\Phi)({\bf k}, t_{dec})&&\nonumber \\
 +\int_{t_{dec}}^{t_0}(\dot{\Psi}-\dot{\Phi})({\bf k},t)
 e^{-i({\bf k} \cdot {\bf n})t} dt] ~. 
\label{dTk} 
\end{eqnarray}

The first term in Eq.~(\ref{dTk}) describes the intrinsic inhomogeneities
on the surface of the last scattering due to acoustic oscillations
prior to decoupling. It also contains  contributions 
to the geometrical perturbations \cite{ram}. The second term describes 
the relative motions of emitter and 
observer. This is the Doppler contribution to the 
CMB anisotropies. It appears on the same angular scale as the
acoustic term and we  denote the sum of the acoustic
and Doppler contributions by  ``acoustic peaks''. The 
last two terms are due to the inhomogeneities in the spacetime 
geometry; the first contribution determines the change in the photon
energy due to the difference of the gravitational
potential at the position of emitter and observer. Together with the part
contained in $D_g^{(r)}$ they represent the ``ordinary'' Sachs-Wolfe 
effect. The second term accounts for red-shifting or blue-shifting
caused by the time dependence of the gravitational field along the 
path of the photon (Integrated Sachs-Wolfe (ISW) effect). The sum of
the two terms is the full  Sachs-Wolfe contribution (SW).

On angular scales  
$0.1^\circ\stackrel{<}{\sim}  \theta\stackrel{<}{\sim}  2^\circ$, 
the main contribution to the CMB anisotropies comes from the acoustic peaks,
while the SW effect is  dominant  on large angular
scales. For topological defects,  the gravitational contribution is
mainly due to the ISW. The ``ordinary'' Sachs Wolfe term even has 
the wrong spectrum, a white noise spectrum instead of
Harrison--Zel'dovich \cite{ruthzhou}.

From  Eq.~(\ref{dTk})
the $C_\ell$'s are  found to be
\begin{equation} 
C_\ell = {2\over \pi} \int 
{<|\Delta_\ell ({\bf k})|^2\rangle  \over (2\ell +1)^2} k^2 dk ~,
\end{equation}
with
\begin{eqnarray}
{\Delta_\ell \over 2\ell +1} &=&j_\ell(kt_0)\left[
{1\over 4}D_g^{(r)}({\bf k},t_{dec}) +(\Psi -\Phi)({\bf k}, t_{dec})\right]
- j_\ell '(kt_0) {\bf V}_r({\bf k},t_{dec})
\nonumber \\
&&+\int_{t_{dec}}^{t_0}(\dot{\Psi}-\dot{\Phi})({\bf k},t')
j_\ell (k(t_0-t')) dt'
\nonumber \\
&=&{1\over 4}D_g^{(r)}({\bf k},t_{dec})j_\ell(kt_0)
- j_\ell '(kt_0){\bf V}_r({\bf k},t_{dec})
\nonumber \\
&& + k\int_{t_{dec}}^{t_0}(\Psi -\Phi)({\bf k},t')j_\ell '(k(t_0-t')) dt'
 ~; 
\label{Dl} 
\end{eqnarray}
$j_\ell$ denotes the spherical Bessel function of order $\ell$ and
$j'_\ell$ stands for its derivative with respect to the argument.

On scales smaller than about $0.1^o$,  the anisotropies
are damped due to the finite thickness of the recombination shell,
as well as by photon diffusion during recombination (Silk damping).
Baryons and photons are very tightly coupled before recombination and
oscillate as one component fluid.
During the process of decoupling, photons slowly diffuse out of
over-dense into under-dense regions. To fully account for this
process, one has to solve the Boltzmann equation (see,
e.g. \cite{RuthReview}). A reasonable approximation can however be
achieved by multiplying the  $\Delta_\ell$  with an exponential damping 
envelope ${\cal D}_{\gamma}(k)$ which is given in Ref. \cite{sh}.

We now discuss the calculation of the $C_\ell$'s for 
perturbations with seeds. Since the background contribution of the
energy momentum tensor of the seeds vanishes, its components
$\Theta_{\mu\nu}$ are gauge invariant perturbation variables.
They can be decomposed into scalar, vector and tensor
contributions.   Here we restrict ourselves to scalar perturbations.  
We express the scalar degrees of freedom of $\Theta_{\mu\nu}$ in terms
of  the gauge invariant perturbation variables
$f_{\rho}, f_p, f_v,f_{\pi}$, which parameterize the energy density, pressure,
scalar velocity potential and anisotropic stress potential of seeds, 
respectively (see \cite{d90,RuthReview}).
\begin{eqnarray}
\Theta_{00}&=&M^2f_{\rho} \\ \Theta^{(s)}_{i0}&=&M^2f_{v,i} \\
\Theta^{(s)}_{ij}&=&M^2[\{f_p-(1/3)\Delta f_{\pi}\}\gamma_{ij}+ f_{\pi,ij}] ~,
\label{seed}
\end{eqnarray}
where $\Delta$ denotes the Laplacian with respect to the metric $\gamma$ of 
the three space and $M$ is a typical ``mass'', energy scale, of the seeds.
The superscript $^{(s)}$ indicates that only the scalar contribution to 
$\Theta_{i0}$ and $\Theta_{ij}$ is obtained in this way. Numerical 
simulations show that the vector and tensor perturbations make up 
about 20\% of the energy momentum tensor on super--horizon scales 
\cite{ruthzhou}. 
Since we assume that the seeds interact with other matter
components only gravitationally, the seed functions satisfy the
following covariant conservation equations  \cite{RuthReview} 
\begin{eqnarray}
\dot f_{\rho}-\Delta f_v +(\dot a/a)(f_{\rho} + 3 f_p)&=&0\label{cons1}\\
\dot f_v +2(\dot a/a) f_v-f_p-(2/3)\Delta f_{\pi}&=&0~.
\label{cons2}
\end{eqnarray}

We consider the matter content of the universe as a two-fluid system: 
the baryons+radiation plasma, which prior to 
recombination is tightly coupled, and cold dark matter (CDM). Before
recombination, the evolution of the perturbation variables in a
spatially flat background, $\Omega = 1$, is described by \cite{KS}
\begin{eqnarray}
\dot D_g^{(r)}-3(c_r^2-w_r){\dot a\over a}D_g^{(r)}+kV_r(1+w_r)&=&0 \label{Dr}\\
\dot D_g^{(c)}+kV_c&=&0\\
\dot V_r+(1-3c_r^2){\dot a\over a}V_r-k(\Psi-3c_r^2\Phi)-
k{c_r^2\over(1+w_r)}D_g^{(r)}
&=&0 \\ 
\dot V_c+ {\dot a\over a} V_c-k\Psi &=&0 ~,
\label{KSeq} 
\end{eqnarray}
where subscripts $_r$, $_c$ (superscript $^{(r)}$, $^{(c)}$) 
denote the baryon-radiation plasma 
and CDM, respectively; $D$ and $V$ are density and velocity perturbations;
$w=p_r/\rho_r$, $c_s^2=\dot{p}_r/\dot{\rho}_r$ and $\rho = \rho_r + \rho_c$.
The geometrical perturbations $\Psi$ and $\Phi$ can be separated into  
a part coming from standard  matter and radiation (subscript $_m$), and a part 
due to the seeds (subscript $_s$).
\begin{eqnarray} 
\Psi &=& \Psi_{m} +\Psi_s \\
\Phi &=& \Phi_{m} +\Phi_s~,
\end{eqnarray}
where $\Psi_s$ and $\Phi_s$ are determined by the energy momentum tensor of 
the seeds. We find  \cite{RuthReview}
\begin{eqnarray} 
\Phi_{m} &=& {4\pi Ga^2\over k^2}[\rho_r D _g^{(r)}+\rho_c D _g^{(c)}
-3\{\rho_r(1+w_r)+\rho_c\}\Phi\nonumber\\
&& +3{\dot a\over a}k^{-1}\{\rho_r(1+w_r)V_r+\rho_cV_c\}]\\
\Psi_{m} &=&-\Phi_{m} \label{Pm}\\
\Phi_s&=&\epsilon k^{-2}[f_{\rho}+3{\dot a\over a}f_v]\\
\Psi_s&=&-\Phi_s -2\epsilon f_{\pi} ~,
\label{pert} 
\end{eqnarray}
where $\epsilon\equiv 4\pi GM^2$. This parameter has to be small to
validate cosmological perturbation theory. In other words, the mass
$M$ has to be significantly smaller than the Planck mass. For global
scalar fields, it actually turns out that the typical amplitude of 
geometrical perturbations is of the order of $\epsilon$, so that the 
COBE normalization requires $ M\sim 10^{16}$GeV.
In this work, we neglect the contribution of neutrino fluctuations.
Anisotropic stresses in the matter components are explicitly set to
zero, $\Pi_{m}\equiv 0$, which implies Eq. (\ref{Pm}). The
anisotropic stresses in the source, $f_\pi$, can therefore not be
compensated. 
 
To solve the above system of equations, we need to specify initial 
conditions. For a given scale $k$, we choose the initial time $t_{in}$
early enough, such that the perturbations are
 super-horizon and the universe is radiation dominated at $t_{in}$.
We set $x=kt$ and denote by a prime the derivative w.r.t $x$. The
super-horizon limit is thus the limit $x\ll 1$. Choosing
$(x,k)$ as independent variables, the
perturbation equations reduce in this  limit to
\begin{eqnarray}
D_g^{(r)'} +{4\over 3}V_r&=&0\label{Dri}\\
D_g^{(c)'} +V_c&=&0\label{Dci}\\
V_r' +(\Phi-\Psi)-{1\over 4}D _g^{(r)}&=&0\label{Vri}\\
V_c'+V_c/x-\Psi&=&0\label{Vci}\\
\Psi_{m} +\Phi_{m}&=&0\label{Pmi}\\
\Phi_s &=& \epsilon k^{-2}(f_{\rho}+3f_v/t)\label{Psi}\\
\Phi &=& {x^2\over 6}\Phi_s + V_r/x \label{Pti}\\
\Psi + \Phi &=& 2\epsilon f_\pi ~. \label{PPi}
\label{init}
\end{eqnarray}
If $f_\rho$, $f_v$ and $f_\pi$ are differentiable in the vicinity of $x=0$, we
can solve the above system exactly. We take the derivative of
Eq. (\ref{Dri}) and replace $V'_r$ with Eq. (\ref{Vri}). Using
Eqs. (\ref{Pti}, \ref{init}, \ref{Psi}) we then find
\be
D_g^{(r)''}+{2\over x}D_g^{(r)'} = 	
	\epsilon(2f_\pi+{x^2\over 3k^2}f_\rho + {x\over k}f_v) ~.
\label{D''} \ee
Differentiability now guarantees that the source term on the right 
hand side of
Eq. (\ref{D''}) is given by $\epsilon A(k)x^\alpha ~ + $ higher orders.
We then obtain to lowest order in $x$
\bea
D_g^{(r)}&=&{\epsilon A(k)\over (\al+2)(\al+3)}x^{\al+2} \\
V_r &=&-{3\epsilon A(k)\over 4(\al+3)}x^{\al+1} \\
\Phi-\Psi &=& {3\epsilon A(k)(\al+1)\over 4(\al+3)} x^\al ~.
\eea
On the other hand, the seed perturbations are of the order of
\be
\Phi_s~,~\Psi_s \propto\epsilon A(k)x^{\al-2} \gg \Phi~,~ \Psi ~, 
\ee
if  $f_\pi\lsim (x^2/3k^2)f_\rho+(x/k)f_v$. For scaling sources, we
shall see that these two terms are of the same order of magnitude. In other
words, $\Phi\ll \Phi_s$ and, if $f_\pi$ is not extremely large, $\Psi\ll
\Psi_s$ on super-horizon scales. 
The main reason for this finding certainly lies in choosing the correct 
initial conditions which have to vanish in the absence of  sources. 
We could always 
add a homogeneous contribution to $D_g^{(r)}$ which would destroy 
this behaviour.  We consider this choice of initial conditions as the
most natural way to obtain compensation:
the presence of matter and radiation reduces the Bardeen potentials on
super-horizon scales by a factor $x^2$.  
Only the contribution 
$\Phi +\Psi=-2\epsilon f_{\pi}$, which is due to anisotropic stresses, 
cannot be compensated by matter and radiation.  
If $f_{\pi} \neq 0$, there is compensation
 provided $f_{\pi} \le \max(f_{\rho}t^2~,~f_vt)$.

We therefore conclude that seeds, which are uncorrelated on super-horizon
scales, are compensated by the presence of matter and radiation,
where we define compensation as the suppression of the total Bardeen
potentials by a factor $x^2$ with respect to $\Phi_s, \Psi_s$.
In this sense, the type of seed perturbations discussed here are
nearly isocurvature fluctuations. 

Within the context of scaling sources (seed functions with white noise
spectra), the basic ingredient which leads to compensation, is
not the absence of perturbations on very large scales or on very early
times, but the fact that we only consider the particular solution of
the second order differential equation for the perturbation variable 
$D_g^{(r)}$. Clearly, a homogeneous contribution to $D_g^{(r)}$ can
destroy this finding. In our case, perturbations are induced by the
presence of the seeds, and therefore these initial conditions are the most
physical ones.

This result is important, since compensation
has usually been understood either as a consequence of the integral
constraint \cite{Traschen,VS,joao} or as a consequence of
causality of the source perturbations\cite{HSW}.
In our work compensation arises naturally  for scaling sources and it
can be generalized to sources with arbitrary spectra which satisfy
\be f_{\pi} \le \max(f_{\rho}t^2~,~f_vt) ~. \label{condcomp} \ee
Clearly for $\Pi=0$, which is the most natural assumption for
non-relativistic cosmic fluids, $f_\pi$ has to be small since it
cannot be compensated, $\Phi,\Psi \gsim {\cal O}(\ep f_\pi)$. 
If $f_\pi$ is larger than the limit given in Eq.~(\ref{condcomp}), then
$\Pi\neq 0$  is a necessary but not sufficient condition for
compensation to occur. 
We know of one example, namely relativistic
collisionless particles, where compensation can take place for
certain choices of 
$f_\pi$, due to the presence of anisotropic stresses $\Pi$ (see appendix). 
Collisionless particles are special in that they interact with
each other and with the seeds only through gravity.
In general, if particle interactions other than
gravity determine $\Pi$, we do not expect compensation, since by
definition the seeds interact with the cosmic fluid only through gravity.
\vspace{0.5cm}

\section{CMB anisotropies induced by scaling sources}
\vspace{0.5cm}
We now restrict our study to scaling  seeds.
We first discuss as motivating example 
global scalar fields which, depending on their number of degrees of
freedom, can lead to global topological defects 
 during a symmetry breaking phase transition in the early 
universe \cite{kibble}. We shall, however, only make use of the
general behaviour of the seed functions $f_{\bullet}$ which we call
``scaling''. In the absence of any intrinsic length scale other than
the cosmic horizon, this is the behaviour which the seed functions
assume by dimensional reasons.

We consider an $N$-component scalar field with potential
$V=\lambda(\phi^2-\eta^2)^2$. In the $\sigma$-model approximation,
the equation of motion for $\phi$ can be expressed solely in the
dimensionless variable $\beta=\phi/\eta$ \cite{ruthzhou}. In terms of
$\beta$ the energy momentum tensor of the scalar field is given by
\begin{equation}
\Theta_{\mu\nu}=\eta^2\left(\beta_{,\mu}\beta_{,\nu}-{1\over 2}
g_{\mu\nu}\beta_{,\lambda}^{,\lambda}\right)~.
\end{equation}
Defining $M^2\equiv \eta^2$, the functions $f_{\bullet}$ result in
\begin{eqnarray}
f_{\rho}(\bf{k})&=&{1\over 2}{\cal F}\left[ \dot {\beta} ^2 +
(\nabla\beta)^2)\right]\nonumber\\
f_p(\bf{k})&=&{1\over 2}{\cal F}\left[\dot {\beta} ^2-{1\over 3}(\nabla\beta)^2)\right]
\nonumber\\
f_v(\bf{k})&=&-{i\over k^2}k^j{\cal F}\left[\dot{\beta}\beta,_j\right]
 \nonumber\\
f_{\pi}(\bf{k})&=& -{3 \over 2k^4}k_ik_j{\cal F}
\left[\beta,_i \beta,_j -{1\over 3}
\delta_{ij}(\nabla \beta)^2\right]~,
\label{f1}
\end{eqnarray}
where ${\cal F}[g]$ denotes the Fourier transform of $g$, 
defined by $g(k)=V^{-1/2}\int e^{ikx}g(x)d^3x$.
In what follows, the Fourier transform of the seed functions
${\cal F}[f_{\bullet}]$, will
be denoted simply by $f_{\bullet}$.

On super-horizon scales $\beta$ and $\dot{\beta}$ are assigned random 
initial values, so they have white noise spectra initially.
$\nabla \beta$ clearly has a $k^2$-spectrum. However, using the 
convolution theorem one finds that $(\nabla \beta)^2$ has a white
noise spectrum.  Therefore, both
$f_{\rho}$ and $f_p$ have  white noise spectra on super-horizon 
scales.  From the above expressions for $f_v$ we find
\[ f_v={(2\pi)^3k^j\over\sqrt{V} k^2}\int d^3q\dot{\beta}({\bf q})\beta({\bf
	q-k})(q-k)_j ~.\]

Expanding this expression in lowest order in $k$ using that $\beta$ and
$\dot{\beta}$ have white noise spectra, we find that the term of order
$k^0$ in the integral vanishes and the lowest order contribution to
the integral is
linear in $k_j$, so that $f_v$ also has a white noise spectrum on
super-horizon scales. By similar arguments one can deduce that
$f_\pi$ has a white noise spectrum on super-horizon scales.

The dimensions of $f_{\rho}$ and $f_p$ in physical space are
(length)$^{-2}$,
therefore in k-space,  $f_{\rho}$ and $f_p$  have dimensions
(length)$^{-1/2}$.
Since on super-horizon scales ($kt\ll 1$), these functions have white
noise
spectra, they must behave as $1/\sqrt t$.
The corresponding arguments lead to a super-horizon behaviour for
$f_v \propto \sqrt t$ and $f_{\pi} \propto t^{3/2}$. So 
the power spectra of the seed functions behave like
\begin{eqnarray}
\langle|f_{\rho}|^2\rangle &=&A_1^2~t^{-1}~F_1(x)\nonumber\\
\langle|f_p|^2\rangle &=&A_2^2~t^{-1}~F_2(x)\nonumber\\
\langle|f_v|^2\rangle &=&A_3^2~t~F_3(x)\nonumber\\
\langle|f_{\pi}|^2\rangle &=&A_4^2~t^3~F_4(x)~,
\label{f2}
\end{eqnarray}
where we choose the dimensionless constants $A_i$ to be positive and such that
$F_i(0)=1$. The power spectra of the functions $f_{\bullet}$ do not 
depend on the direction of {\bf k}, thus the $F_i$'s are even 
functions of $x=kt$. Furthermore,
since the energy momentum tensor of the source decays inside the
horizon, we know that $F_i\rightarrow 0$ for $x\rightarrow
\infty$. This behaviour of $f_\rho$, $f_p$ and $f_v$ has also been
found by numerical simulations.

The temporal behaviour of $f_\rho$ and $f_p$ can also be understood
from the following argument: the $k=0$ component of $f_\rho$ just
corresponds to the average energy density multiplied by $\sqrt{V}$ and
is thus proportional to $ V^{1/2}/t^2$. On super-horizon scales,    
$f_\rho(k)$ is white noise superimposed on this average.
The number $N$ of independent patches in $V$ is 
$V/t^3$ and hence the amplitude of  $f_\rho(k)$ is proportional to 
$V^{-1/2}/(t^2N^{-1/2})\propto t^{-1/2}$. The same arguments hold for $f_p$.
From numerical simulations \cite{ruthzhou} for $\pi_3$ defects, 
global textures, one finds that the average of $\dot{\beta }^2$ over a shell 
of radius $k$ can be modeled on super-horizon scales by  
\begin{equation}
\langle|\dot{\beta}^2|^2\rangle (k,t) \sim {2 \over t }.
\end{equation}

We define a seed to be scaling if the power spectra of the seed
functions behave as in Eq.~(\ref{f2}). We expect this scaling behaviour
to be 
valid not only for global scalar fields, but also for (local) cosmic strings.
However, since the only decay mechanism for cosmic strings is through
emission of gravitational radiation, we expect the functions $F_i$ to
decay slower on sub-horizon scales, than in the case of global fields,
which decay very efficiently into Goldstone bosons.  

The system given by Eq.~(\ref{Dri}) to  (\ref{init}) can be solved 
analytically if the stochastic
variables $f_{\bullet}$ are replaced by the square root of their power spectra.
The results are thus to be taken with a grain of salt. But we
believe that the r.h.s. of the following equations are good approximations
to the square roots of the power spectra of the corresponding
stochastic variables on the l.h.s., since, as we argue below, this 
coherence assumption does not significantly influence the results on 
super--horizon scales. Inserting the square roots of Eqs.~(\ref{f2}) in 
the system (\ref{Dri}) to (\ref{PPi}), one finds in the limit $x\ll 1$
\begin{eqnarray}
\Phi_s&=&\epsilon(A_1+3A_3)k^{-3/2} x^{-1/2} \label{Pscorsuper}\\
\Psi_s&=&-\epsilon(A_1+3A_3+2A_4x^2)k^{-3/2} x^{-1/2}\\
D_g^{(r)}&=&{16\over 189}\epsilon (2A_4 +{1\over 3}A_1 +A_3) 
k^{-3/2} x^{7/2}  \label{Dcorsuper}\\
D_g^{(c)}&=&{2\over 63} \epsilon (4 A_4 +{5\over 21}A_1+{5\over 7}A_3) 
k^{-3/2}x^{7/2} \label{Dccorsuper}\\
V_r&=&-{2\over 9}\epsilon ({1\over 3}A_1+A_3+2A_4)k^{-3/2}x^{5/2}\\
V_c&=&-{\epsilon \over 63}(28A_4 +{5 \over 3}A_1+5A_3)k^{-3/ 2}x^{5/2}\\
\Phi&=&{1\over 6}x^2 \Phi_s +{V_r\over x}\\
\Psi&=&-\Phi -2\epsilon A_4k^{-3/2}x^{3/2} ~. \label{Pcorsuper}
\end{eqnarray}
We use these results as initial conditions for the system (\ref{Dr}) to 
(\ref{pert}). Eqs. (\ref{Dcorsuper}) and (\ref{Dccorsuper}) show, that
the perturbations are in general non-adiabatic.

Due to the conservation equations (\ref{cons1}) and (\ref{cons2}), 
the constants $A_i$ are not independent.
Taking the sum of the ensemble averages of the  conservation
 equation Eq. (\ref{cons1}) multiplied by 
$f_{\rho}^{\star}$ (the complex conjugate of $f_{\rho}$), and its complex 
conjugate, we obtain
\begin{eqnarray}
{d\over dt} \langle|f_{\rho}|^2\rangle  +k^2\langle f_{\rho}^{\star} f_v+f_{\rho}f_v^{\star}\rangle 
+2{\dot{a}\over a}\langle|f_{\rho}|^2\rangle &&\nonumber\\ 
+3{\dot{a}\over a}\langle f_{\rho}^{\star} f_p+f_{\rho}f_p^{\star}\rangle &=&0.
\end{eqnarray}
Let us discuss the above equation on super-horizon scales where we can 
neglect the second term.
Since  $ (d/dt) \langle|f_{\rho}|^2\rangle=-\langle|f_{\rho}|^2\rangle /t$, we find that the real 
part of $\langle f_{\rho}^{\star} f_p\rangle $ is negative. Furthermore, Schwarz inequality
leads to
\begin{equation}
|A_2| \ge {1\over 6}|A_1| ~\mbox{ in the radiation era, and } |A_2|\ge 
{1\over 4}|A_1|
 ~\mbox{ in the matter era.} \label{A12}
\end{equation}
Similarly, from Eq. (\ref{cons2}) we conclude that the real part of 
$\langle f_v^{\star}f_p\rangle $ is positive and 
\begin{equation}
|A_2| \ge {5\over2}|A_3| ~\mbox{ in the radiation era, and } |A_2|\ge 
{9\over2}|A_3| 
 ~\mbox{ in the matter era.} \label{A23}
\end{equation}
The equality sign is valid, if and only if
\[ \langle f_{\rho}^{\star} f_p+f_{\rho}f_p^{\star}\rangle  = 
	-2\sqrt{\langle|f_\rho|^2\rangle 
	\langle|f_p|^2\rangle } ~.\]
We call this condition, which requires that $f_\rho$ and $f_p$ are in perfect
phase correlation ``perfect coherence'' between $f_\rho$ and
$f_p$. On super-horizon scales the
spectrum of $k^2f_v$ is much smaller than the spectrum of $f_\rho/t$;
and thus $k^2f_v\langle\langle f_\rho/t$, almost everywhere in the space of
realizations. Hence, on super--horizon scales energy conservation
 (Eq. \ref{cons1}) yields 
\[ f_\rho(t)={a(t_{in})\over a(t)}f_\rho(t_{in})
-{3\over a(t)}\int_{t_{in}}^t\dot{a}(t')f_p(t')dt'~.\]
The question of  coherence between $f_\rho$ and $f_p$ is thus reduced
to the question of unequal time coherence of $f_p$ and $f_\rho$
themselves. Similarly, the coherence between $f_v$ and $f_p$ reduces
to the unequal time coherence of each of these functions. 
We believe that for scaling sources, unequal time coherence is
reasonably well maintained on super-horizon scales and therefore the
equal signs in Eqs. (\ref{A12}) and (\ref{A23}) are probably valid on 
sufficiently large scales. Numerical simulations for global scalar 
fields and the large $N$ limit (see \cite{martin}) support this
hypothesis.

We now  address the effect of  unequal time coherence, $\langle f_i(t)f_j(t')\rangle $,
on the resulting power spectrum $C_\ell$.
To simplify the relevant equations, we neglect here the short matter
dominated period before decoupling and we also neglect baryons, such
that $c_r^2=w_r=1/3$. The dynamical components during the tight
coupling epoch are thus reduced to radiation and seeds.
A WKB solution of the evolution equations then gives 
\begin{eqnarray}
D_g^{(r)}(k,x)&=&
{4\over \sqrt 3}\int_0^x dx' [\Phi(x') -\Psi(x')]~
sin((x-x')/\sqrt{3})\nonumber\\
V_r(k,x)&=&-\int_0^{x} dx'
[\Phi(x')-\Psi(x')]~cos((x-x')/\sqrt{3})~, \label{WKB}
\end{eqnarray}
where we set $\Phi(0)=\Psi(0)=D_g^{(r)}=0$.\\
This actually just reformulates our simplified system of equations in
terms of two integral equations, since the Bardeen potentials $\Phi$
and $\Psi$ are given in terms of $D_g^{(r)},~V_r$ and the source
functions as follows
\bea 
\Phi-\Psi &=& {2\over 6+x^2}({3\over 2}D_g^{(r)} + {6\over x}V_r)
+2\ep( {x^2\over k^2(6+x^2)}f_\rho +{3x\over k(6+x^2)}f_v +f_\pi) 
\nonumber \\
 &=&  {2\over 6+x^2}({3\over 2}D_g^{(r)} + {6\over x}V_r) +{2x^2\over
6+x^2}\Phi_s +2\ep f_\pi ~.
\eea
As we have seen earlier, it is a very bad approximation to replace $\Phi$ 
and $\Psi$ by the corresponding source potentials $\Phi_s$ and $\Psi_s$, 
since this does not take care of the compensation.

Applying the Hu and Sugiyama formalism \cite{husu} for topological
defects \cite{joao}, we obtain within our approximation ($\Om_b=0$,
purely radiation dominated)
\begin{eqnarray}
{\Delta_\ell(k) \over 2\ell +1} &=&\int_0^{x_{dec}}dx 
[\Phi(x)-\Psi(x)]\{{j_\ell(x_0-x_{dec})\over \sqrt{3}}
\sin({x_{dec}-x\over\sqrt{3}})  \nonumber\\
&& -j'_\ell(x_0-x_{dec})\cos({x_{dec}-x\over \sqrt{3}})\}
\nonumber \\
&&+[\Psi (x_{dec}) -\Phi (x_{dec})] j_\ell (x_0-x_{dec})\nonumber\\
&&+\int_{x_{dec}}^{x_0}dx [\Psi'(x)-\Phi'(x)]~j_\ell(x_0-x),
\label{coher1}
\end{eqnarray}
where $x_{dec}=kt_{dec}$, $x_0=kt_0$. The times $t_{dec}$ and  $t_0$ denote 
the time of decoupling and today respectively. The somewhat more involved
formula which takes into account the presence of baryons and CDM, can
be found in \cite{joao}.

Assuming coherence, the power spectrum $C_\ell$ can be calculated by squaring 
$\Delta_\ell / (2\ell +1)$ obtained from  
Eq.~(\ref{coher1}), where each variable is replaced by the square root of its
power spectrum. On the other hand, for totally incoherent perturbations, one
should instead use \cite{joao}
\begin{eqnarray}
C_\ell&=&\int dkk^3 \int_0^{x_{dec}} dx  P_r(\Phi-\Psi)
\{{j_\ell(x_0-x_{dec})\over \sqrt{3}}
\sin({x_{dec}-x\over\sqrt{3}}) \nonumber \\
&&  -j'_\ell(x_0-x_{dec})\cos({x_{dec}-x\over\sqrt{3}})\}^2
\nonumber \\
&&+\int dk k^2 [P(\Psi -\Phi)(t_{dec})]~j_\ell^2(x_0-x_{dec})\nonumber\\
&&+\int dk k^3 [j'_\ell(x_0-x_{dec}) j_\ell(x_0-x_{dec})
[P_r(\Psi-\Phi)](x_{dec})]\nonumber\\
&&+\int dk~\int_{x_{dec}}^{t_0}  dx k^2P_r( \Psi'-\Phi')
j_\ell^2(x_0-x)~,
\end{eqnarray}
where $P(X)\equiv \langle|X|^2\rangle (k,t)$ denotes the power spectrum of the
variable $X$ and $P_r(X)$ is the power spectrum of $X$ integrated over
a short time period $\Delta t$ (see \cite{joao}).

We now want to illustrate the difference of the two approaches in our 
simplified pure radiation model. For pure radiation we can derive the
following second order equation for $D_g^{(r)}$:
\begin{eqnarray} 
D_g^{(r)''} +{12\over(6+x^2)x}D_g^{(r)'}-{2-x^2/3\over
	6+x^2}D_g^{(r)} &=&{8\ep\over 3}\left[ 
	{x^2\over k^2(6+x^2)}f_\rho +{3x\over k(6+x^2)}f_v +f_\pi\right]
	\nonumber\\
	&=& {8\over 3}( \epsilon f_{\pi} +{x^2\over 6+x^2} \Phi_s) ~,
  \label{Drad} \end{eqnarray}
with general solution
\be
D_g^{(r)}={8\over 3} \ep \int _0 ^x  f(x') G(x,x') dx' ~,
\ee
where $f(x)\equiv  f_{\pi}(x)+ (x^2/6\epsilon) \Phi_s/$ 
and $G(x,x')$ denotes the Green's function
\be
G(x,x')={\sqrt 3 x'\over (6+x^{'2})x}\left[(12+xx')\sin\left({x-x'\over 
\sqrt 3}\right) 
   +2\sqrt 3 (x+x')\cos \left({x-x'\over\sqrt 3}\right)\right]~.
\ee
The power spectrum of $D_g^{(r)}$ is therefore
\be
\langle|D_g^{(r)}|^2\rangle  = {64\over 9}\epsilon ^2\int _0 ^x\int _0 ^x dx' dx''
\langle f(x')f^{\star}(x'')\rangle  G(x,x') G(x,x'')~.
\label{coherence1}
\ee
Assuming total coherence, Eq.~(\ref{coherence1}) takes the form
\begin{eqnarray}
\langle|D_g^{(r)}|^2\rangle  &=&[{8\over 3}\ep
\int _0 ^x dx' \sqrt {\langle|f(x')|^2\rangle } G(x,x') ]^2~.
\end{eqnarray}
On the other hand, assuming complete decoherence,
\be
\langle f(x)f^{\star}(x')\rangle =\delta(x-x')\int_x^{x+\Delta x}dx \langle|f(x)|^2\rangle ~, 
\ee
leads to the power spectrum
\be
\langle|D_g^{(r)}|^2\rangle  ={64\over 9}\epsilon ^2r \int _0 ^x dx' 
\langle|f(x')|^2\rangle  G^2(x,x')~, \label{wron}
\ee
where we have chosen $\Delta x=rx$.
We further assume also complete decoherence between  different source 
functions,
\be
\langle f_{\pi}f_v\rangle =\langle f_{\pi}f_{\rho}\rangle =\langle f_vf_{\rho}\rangle =0~.
\ee
In Figs.~1a and 1b we plot $|D_g^{(r)}|^2 k^3$ versus $k t_{dec}$
under the assumption of total coherence and complete decoherence 
respectively. The role of the coherence assumption on the characteristics 
of the power spectrum is shown in Fig.~2. Clearly, complete decoherence
shifts the first acoustic peak to smaller angular scales, and
reduces substantially its height. Furthermore, 
secondary peaks are completely washed out. 

A realistic defect model will always lay somewhere between these two
extremes. We suppose however, motivated by numerical simulations of
textures and the large $N$ limit, that the texture example is closer to
the completely coherent case. In the next section we thus restrict
ourselves to perfect coherence.

\vspace{0.5cm}
\section{Numerical examples}
\vspace{0.5cm}
In this section we study how the characteristics of the acoustic peaks depend 
on the values of the dimensionless constants $A_i$ and the form 
of the functions $F_i$, which determine the power spectra of the seed functions
(see Eq.~(\ref{f2})).  A  crucial question is  whether there is
a set of parameters for which the position and amplitude of the primary 
acoustic peak are similar to those predicted by an adiabatic  inflationary
model. 

As we discussed earlier, the functions $F_i$ are normalized such 
that $F_i(0)=1$, and 
$F_i\rightarrow 0$ for $x\rightarrow \infty$. Numerical simulations for 
global textures \cite{ruthzhou} suggest that in the case of global scalar 
fields, the functions $F_i$  have power law decay.  However, for generic 
scaling sources, one could also consider the case of exponential decay. 
As we shall show, the form of these functions affects the features of the power
spectrum significantly. In general, we find that if $F_i$ have an 
exponential decay, the position of the primary peak is within the
range  predicted by adiabatic inflationary 
models, at $\ell \sim 220$.  On the other hand, if $F_i$ have a power law
decay, as it seems to be for global topological defects, the position of
the first acoustic peak is clearly shifted to smaller angular scales, at 
around $\ell \sim 300$ to $400$.  The amplitude of fluctuations
decreases by up to a factor of 500 if we choose an exponential 
 decay law for the seed functions. This is due to the fact that
 in this case the decay is very fast and erases
almost all substantial seed contribution. 
The sensitivity of the overall amplitude of CMB perturbations on the
 parameters is extremely important especially if one wants to rule out defect
 models with biasing arguments!

In figures~3 to~8, we show the resulting power spectra for different
set of dimensionless constants $A_i$, with the seed functions having either
exponential or power law decay. The dashed line indicates the SW contribution,
the dashed-dotted line is the contribution from the acoustic peaks,
and their sum is drawn as solid line. 

In Fig.~3a we choose an exponential 
decay for all the seed functions, $F_i(x)= \exp(-x^2)$, and a set of
constants $A_1=3, A_3=-0.6, A_4=0$.  The position of the primary peak is at
$\ell \sim 200$, while the relevant height of the first peak with respect
to the SW plateau is  $\sim 25$.  Using the same set of constants $A_i$,
however choosing a power law decay for the seed functions, 
$F_i(x)=[1.+ (1./(2 \pi)^2) x^2 ]^{-1}$, we see in Fig.~3b that the 
peak is displaced to smaller angular scales, at $\sim 330$, while the
relative amplitude of the acoustic peak with respect to the SW plateau remains
$\sim 25$. Also the features of the secondary peaks are different.  
While in Fig.~3a
the second and third peaks have almost the same height, in Fig.~3b the
second peak has almost completely disappeared. 
In both cases, the spectral index of the plateau, in the range 
 $\ell \sim 2-20$ is 
$n\sim 1$, consistent with observations. 

Now, selecting a slightly different set of dimensionless constants, we see that
the predicted power spectrum is very different. In Fig.~4 we show the
power spectrum for $A_1=3, A_3=-0.7, A_4=0$, and the same power law decay
for the functions $F_i$  as in Fig.~3b.  The primary peak is again at rather
large angular scales, $\sim 350$, but the height of the peak is different,
and the spectral index clearly deviates from 1. 

A very interesting case is displayed in Fig.~5a, where we see that both the
position and the amplitude of the first acoustic peak, agreed with those
predicted by a generic inflationary model. Here, $A_1=3, A_3=1, A_4=2$,
and $F_i(x)=\exp(-x^2)$.  The primary peak is at 
$\ell \sim 200$ and its relative amplitude is at $\sim 4$. The second peak is 
almost completely washed out and the spectral index in the range 
$\ell \sim 2-20$ is very close to 1. This power spectrum, where 
perturbations are generated by scaling seeds, is quite 
similar to one resulting from an adiabatic inflationary model. 
Considering the same set of parameters and a somewhat slower
 exponential decay for the functions
$F_i$  given by $F_i(x)= exp(-0.5 x^2)$, we find (Fig.~5b)
that both, the position and relative amplitude of the first peak 
with respect to the SW plateau, remain the same as in
Fig.~5a, whereas the Sachs Wolfe plateau is somewhat prolonged.
This simple example shows that it may well be possible to
``manufacture'' inflationary spectra by a suitable choice of seed
functions. A point which has already been realized in
Ref.~\cite{neil}. 
It is thus extremely important to further constrain the seed
functions of defect models by numerical simulations and/or the large $N$ limit.
It may well be that the requirement of power law decay of the seed
functions, excludes the inflationary position of the first acoustic
peak. With the same parameters $A_i$ and a power law decay 
$F_i(x)=[1+ (1/(2 \pi)^2) x^2 ]^{-1}$, the primary peak is at $\ell \sim
320$, while the relative height of the peak is about $ 8$, Fig.~5c.

We have also considered the values for the parameters $A_i$ which are 
suggested by the conservation equations and perfect coherence,
choosing a power law decay,
 $F_i(x)=[1+ (1/(2 \pi)^2) x^2 ]^{-1}$.
We find that the sign of $A_3$ does not affect the
features of the power spectrum.
In Fig.~6a  $A_1=3, A_3=-1/6, A_4=1/(2(2\pi)^2)$, while in Fig.~6b
$A_1=3, A_3=1/6, A_4=1/(2(2\pi)^2)$. In both these cases, the peak is at
$\ell \sim 300$, and  its amplitude is $\sim 8$.  

Finally, to illustrate the variety of results which can be obtained by
parameter variation within a simple family of seed functions, we show 
a rather extreme case in Fig.~7, where $A_1=3, A_3=-1., A_4=0$ 
and $F_i(x)=exp(-x^2)$.
Here we see  no acoustic peaks at all. However, this is a 
rather particular case, since with this choice of parameters, all the
variables $\Phi_s, \Psi_s, D_g^{(r)}, D_g^{(c)}, V_r$ and $V_c$
vanish initially (see Eqs. (\ref{Pscorsuper}) to (\ref{Pcorsuper})).

To analyze the dependence of the characteristics of the power spectrum
on the seed functions in a somewhat more systematic way, we have calculated
the $C_\ell$'s for a grid of values $-A_1\le A_3,A_4\le A_1$ with
spacing $0.2A_1$ and fixed functions $F_i=1/(1+(x/(2 \pi))^2)$. We
fitted the resulting $C_\ell$ for $2\le \ell\le 20$ to the simple
power law behaviour arising in inflationary models, $C_\ell\propto
\Ga(\ell+(n-1)/2)/\Ga(\ell+(5-n)/2)$. We find that for $|A_4|\ge
0.4A_1$, the $\chi^2$ of the fit is unacceptably high: $\chi^2 \sim 3$
to $4$ for  $|A_4|=0.4A_1$, and more than 10 for even larger
anisotropic stresses. We allow for a relative error of 0.05.
However, $\chi^2$ depends only weakly on the
value of $A_3$ (see Fig.~8). We therefore restrict the parameter range
for $A_4$  to $-0.3A_1\le A_4\le 0.3A_1$. The spectral index is in
good agreement with observations, $1\le n\le 1.4$ (see Figs.~9a 
and ~9b). 

We find positions of the first acoustic peak in the range 
$260\le \ell_{peak}\le 520$. For the choice of  seed functions with 
power law decay on sub-horizon scales (which is also indicated from numerical 
simulations and from the large $N$ limit), we never obtain the peak at the 
adiabatic inflationary position of $\sim 220$, and values $l_{peak}
<300$ are only found for very small $A_4$ (see Fig.~10). 

We define 
the quantity $\ell_{peak}(\ell_{peak}+1)C_{\ell_{peak}}/(110C_{10})$ as a
measure for the height of the acoustic peak. This quantity is very model 
dependent and assumes, within the small class of models investigated in our 
parameter study, all values between $0.1$ (for $A_4=0$ and $A_3=A_1$,
 i.e., virtually no discernible peak) and $11$ (see Figs. 11a and 11b). 
For fixed $A_3$, the peak height is a steeply raising function of $|A_4|$.
Only values $|A_4| < 0.1A_1$ lead to peak heights below 6. 

It is interesting to note, that also the absolute amplitude of the 
spectrum is sensitively depending on the ratios $A_4/A_1$ and $A_3/A_1$.
For $A_4=0$ the amplitude $110C_{10}$ varies from $0.002\epsilon^2A_1^2$ 
for $A_3=-0.2A_1$ to $0.8\epsilon^2A_1^2$ for $A_3=A_1$. If $A_4\neq 0$, the 
amplitude does not depend very strongly on $A_3$ and grows from 
$\sim 0.4\epsilon^2A_1^2$ for $|A_4|=0.1A_1$ to $\sim 4\epsilon^2A_1^2$ 
for $|A_4|=0.3A_1$ (see fig.~12). This finding is important for 
 the biasing problem of structure formation.
Sometimes, defect models have been 
claimed to be ruled out, since they would not lead to large 
enough matter density fluctuations, if normalized to the COBE experiment 
on very large scales. This normalization fixes the only free parameter
of a given model, namely the symmetry breaking scale and therefore $\ep$.
In our work, we have seen that the  Sachs Wolfe fluctuations in 
the CMB are largely governed by  $A_4$, the amplitude of anisotropic 
stresses. 
The density fluctuations in the dark matter, however, are induced by 
$\dot{\phi}^2$ alone (see, e.g. \cite{d90}), which is determined
entirely by $A_1$ and $A_2$.
Defect models, with somewhat small anisotropic stresses,
e.g., $A_4<0.05A_1$, which are actually quite natural, but difficult
to resolve numerically, may explain the different bias factors
 obtained from numerical simulations in \cite{pst,ruthzhou}.

\vspace{0.5cm}
\section{Conclusions}
\vspace{0.5cm}
In this paper we analyzed with some generality the CMB anisotropies
induced in models with scaling sources for an $\Om=1$, cold dark
matter cosmological model.

Within the framework of gauge invariant perturbation theory it turns
out that compensation is a consequence of ``natural'' initial conditions.
By ``natural'' we mean that we only consider that part of the solution
induced by the source itself and do not add an arbitrary homogeneous
contribution. In this case, we have found that the total Bardeen
potentials are reduced by a factor $x^2=(kt)^2$ with respect to the
potentials generated by the source alone. One may think at first sight
that such a result is unphysical, acausal, however it just reflects
that also the initial condition of a perfect Friedmann universe is
acausal. 

Even restricting ourselves to the case of scaling sources, we found
that the resulting power spectrum depends significantly on the model 
parameters. In particular, if the seed functions decay exponentially,
the position of the first acoustic peak is at $\ell \sim 220$ as in
inflationary models. Adjusting the amplitude of the seed functions, we
can also obtain a peak height consistent with inflationary perturbations.

On the other hand, if the seed functions have a power law decay, as
numerical simulations of global textures \cite{ruthzhou} as well as
the large $N$ limit \cite{martin} indicate, the position of the first
peak is within the range $260\le \ell_{peak}\le 500$. Its amplitude
depends sensitively on the parameters of the seed functions which, for
a specific model, have to be determined by involved numerical
simulations. In our analysis we encountered amplitudes in the range
$0.1\le \ell_{peak}(\ell_{peak}+1)C_{\ell}^{peak}/(110C_{10})\le 25$.

We also found that the total amplitude of CMB anisotropies produced
depends strongly on the amplitude of anisotropic stresses of the
seed. Whereas, the source term leading to CDM density 
fluctuations is given by $f_\rho+3f_p$, i.e, determined by $A_1+3A_2$ on
large scales. Therefore, the relation between the COBE normalization 
of the model and the bias factor depends sensitively on the ratio
$A_4/(A_1+3A_2)$, which may depend on details of the model.

These results are obtained under the assumption of perfect
coherence. This hypothesis seems reasonable for global scalar fields
as also the large $N$ limit indicates \cite{martin}. On the other
hand, assuming complete decoherence, the position of the first peak is
shifted to smaller angular scales, its amplitude is reduced and
secondary peaks are washed out. A realistic situation may lay
somewhere between the two extremes.

The  examples with power law seed functions which we discussed in 
this paper were motivated by numerical
simulations of global textures with vacuum manifold ${\bf S}^3$. Apart
from the scaling behaviour on very large and very small scales, which
should be the same for all global defects, we do not know to what
extend the seed functions depend on this particular choice.  

\vspace{0.5cm}
{\Large\bf Acknowledgment}\hspace{0.5cm} It is a pleasure to thank 
Alejandro Gangui and Martin Kunz for helpful
suggestions. We also thank Nathalie Deruelle, Maurizio Gasperini and
Gabriele Veneziano for stimulating
discussions. This work is partially supported by the Swiss
NSF. M.S. acknowledges financial support from the Tomalla foundation.
\vspace{0.5cm}

\vspace{1.5cm}
\appendix
\setcounter{equation}{0}
\renewcommand{\theequation}{A\arabic{equation}}

{\LARGE \bf APPENDIX}

\section{Compensation for relativistic collisionless particles }
\vspace{0.5cm}
In this appendix we  show how compensation arises in the 
case of relativistic collisionless particles. 

We consider a universe dominated by massless ({\em i.e.} relativistic)
collisionless particles, with scalar perturbations induced by
seeds. We assume seeds consisting of massless particles  conformally
coupled to gravity. In this case, the time dependence of the seed
functions is given by $f_\bullet\propto 1/a^2 \propto 1/t^2$ ($t$ denotes
conformal time).

The evolution of perturbations is determined by the
collisionless Bolztmann equation, which reads \cite{RuthReview}
\be
\dd_t{\cal M}(\mu,k,t) +ik\mu {\cal M}=ik\mu[\Phi-\Psi](k)~,
\label{liou}
\ee
where $k$ denotes the wave number and $\mu={\bf n\cdot k}/k$;
{\bf n} stands for the momentum direction of the
relativistic particles. $\cal M$ is a gauge-invariant perturbation
variable for the energy integrated one-particle distribution function,
\[ {\cal M}={\pi\over\rho}\int p^3dp\de f ~.\]
Using the general definition of the energy momentum tensor,
\[ T^{\mu\nu}=\int {d^3p\over p^0}f(p)p^\mu p^\nu  ~,\]
we obtain
\bea
D_g &=& 2\int_{-1}^1{\cal M}d\mu    \label{Dg} \\
V &=& {3i\over 2}\int_{-1}^1{\cal M}\mu d\mu    \label{V}  \\
\Pi &=&3\int_{-1}^1(1-3\mu^2){\cal M}d\mu  ~.  \label{Pi}  \eea

The gravitational perturbation equations on super-horizon scales yield
\bea 
\Phi &=& {1\over 4}D_g+{V\over x} +{1\over 6}x^2\Phi_s \label{PhiA}\\
\Psi &=& -\Phi-2\ep f_\pi -{1\over x^2}\Pi  \label{PsiA} ~,\eea
where $\Phi_s =\ep k^{-2}[f_\rho +(3/t)f_v]$.
Furthermore, the conservation equation, $D'_g=-(4/3)V$, tells us that
$D_g \propto xV$, so that we may neglect the $D_g$ term in Eq.~(\ref{PhiA}).
Inserting Eqs.~(\ref{PhiA}, \ref{PsiA}) in Eq.~(\ref{liou}), we obtain
\be
\dd_y{\cal M}(y,x) +i{\cal M}=i[{2V\over x} +{1\over 3}x^2\Phi_s+2\ep
	f_\pi +{\Pi\over x^2}](k)~,
\ee
where $y=\mu x$.

For $x\ll 1$, and thus  $y\ll 1$, we make the ansatz (see
\cite{Peebles})
\be {\cal M}=x^\beta[c_1y +c_2y^2+{\cal O}(y^3)] \label{ansatz} ~, \ee
where $c_1,~c_2$ and $\beta$ are constants.
Inserting this ansatz in Eqs.~(\ref{Dg}), (\ref{V}) and (\ref{Pi}), we
get
\bea
D_g &=& {4\over 3}c_2x^{\beta+2}  \\
V &=& ic_1x^{\beta+1}  ~~~~~ = -(\beta +2)c_2x^{\beta+1}  \label{Vsup}
\\
\Pi &=& -{8\over 5}c_2x^{\beta+2}~,
\eea
where the second equality in Eq.~(\ref{Vsup}) is obtained from the
energy conservation equation (the zeroth moment of Eq.~(\ref{liou})).
The first moment of Eq.~(\ref{liou}) (momentum conservation)
implies
\be [(\beta+2)(\beta+3)+{8\over 5}]c_2x^\beta = {1\over 3}x^2\Phi_s
+2\ep f_\pi ~. \label{momentum}\ee
If $\ep f_\pi\lsim x^2\Phi_s$, Eq.~(\ref{momentum}) leads to 
${\cal O}(V/x) = {\cal O}(\Pi/x^2) = {\cal O}(x^2\Phi_s) = 
   {\cal O}(x^2\Psi_s)$. 
Thus, ${\cal O}(\Phi)={\cal O}(x^2\Phi_s)$ and 
${\cal O}(\Psi)={\cal O}(x^2\Psi_s)$ which means that we find
compensation. In the other case, $f_\pi \gg x^2\Phi_s$, the right
hand side of Eq.~(\ref{momentum}) behaves like $\ep f_\pi\propto
1/x^2$, implying
$\beta=-2$. Eq.~(\ref{momentum}) then leads to
\[ -\Pi/x^2 ={8\over 5}c_2x^\beta \sim 2\ep f_\pi  ~~~~\mbox{ and }~~~
V\sim 0~.\]
Inserting this in Eqs.~(\ref{PhiA}, \ref{PsiA})  yields
\be \Psi =-\Phi= -{1\over 6}x^2\Phi_s ~,\ee
and thus ${\cal O}(\Phi)= {\cal O}(x^2\Phi_s)$ and ${\cal O}(\Psi)\le 
{\cal O}(x^2\Psi_s)$, leading again to compensation.

We thus have shown that in this  example, compensation is
present even if the anisotropic stresses of the seeds are not
suppressed.

\newpage

\begin{figure}[htb]
\centering
\epsfysize=10.5cm
\epsffile{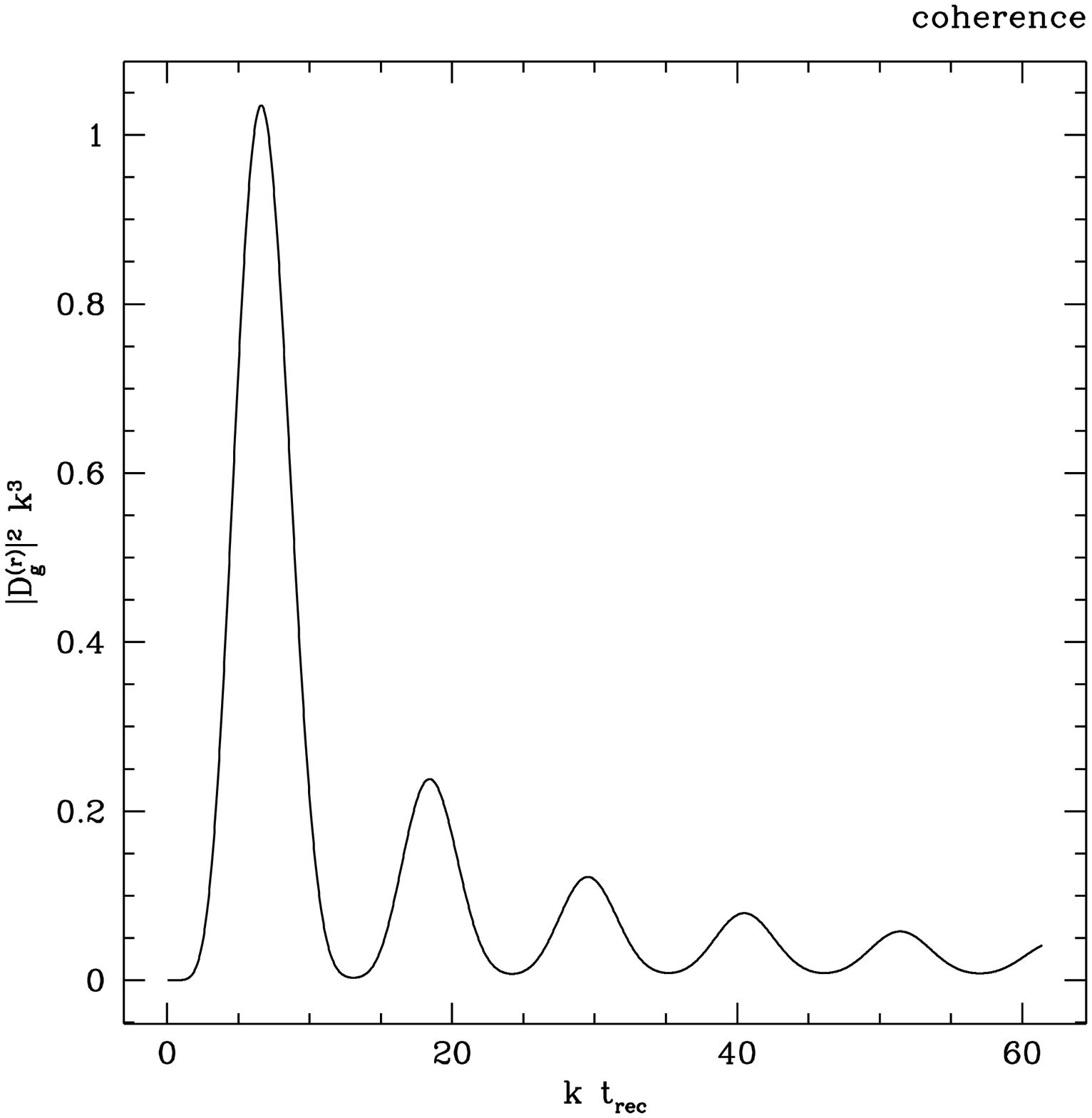}
\epsfysize=10.5cm
\epsffile{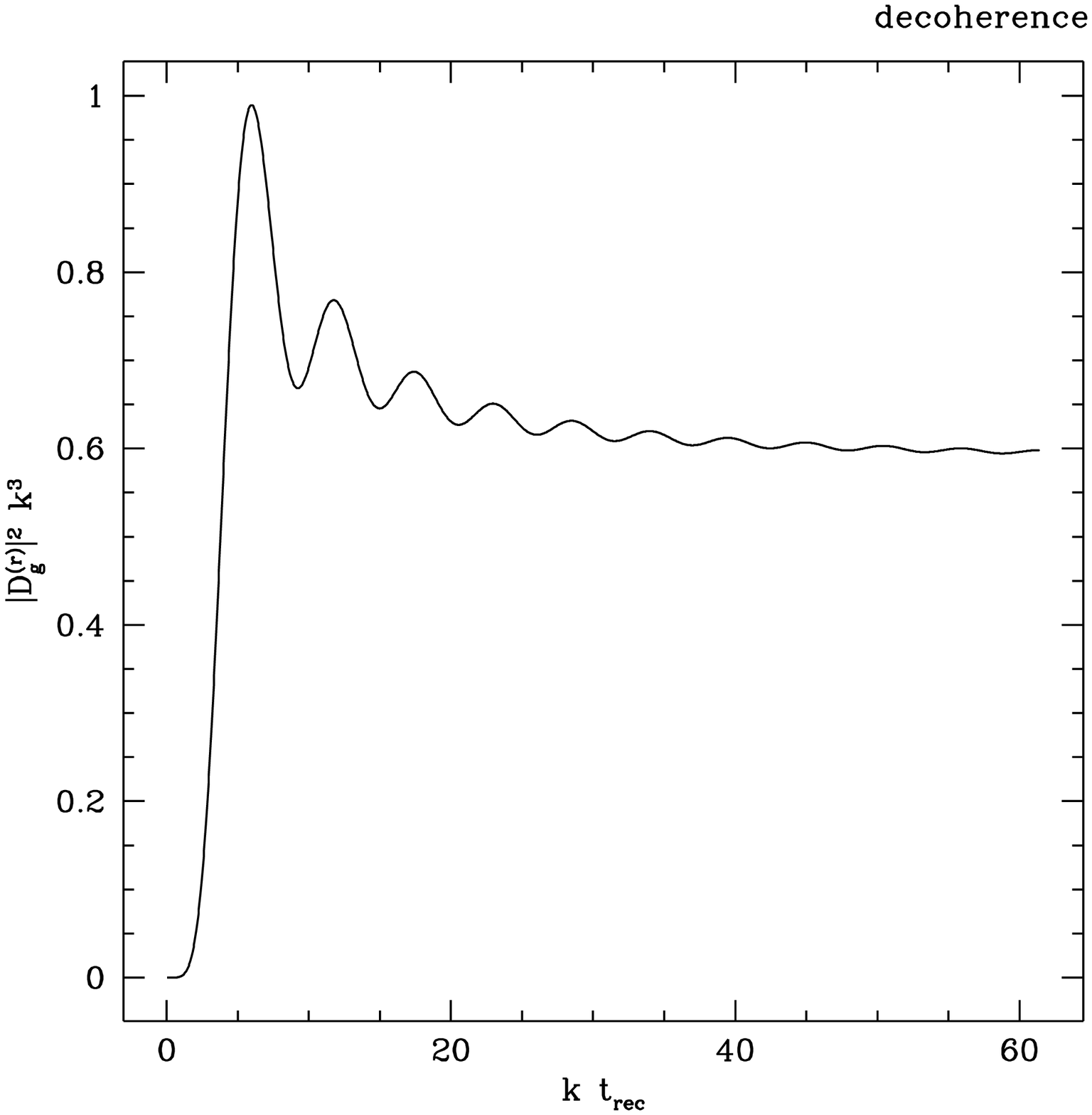}
\caption{The acoustic fluctuations in the photon density spectrum 
are shown for the case
of perfect coherence, (top) and complete decoherence (bottom).}
\end{figure}

\begin{figure}[htb]
\centering
\epsfysize=10.5cm
\epsffile{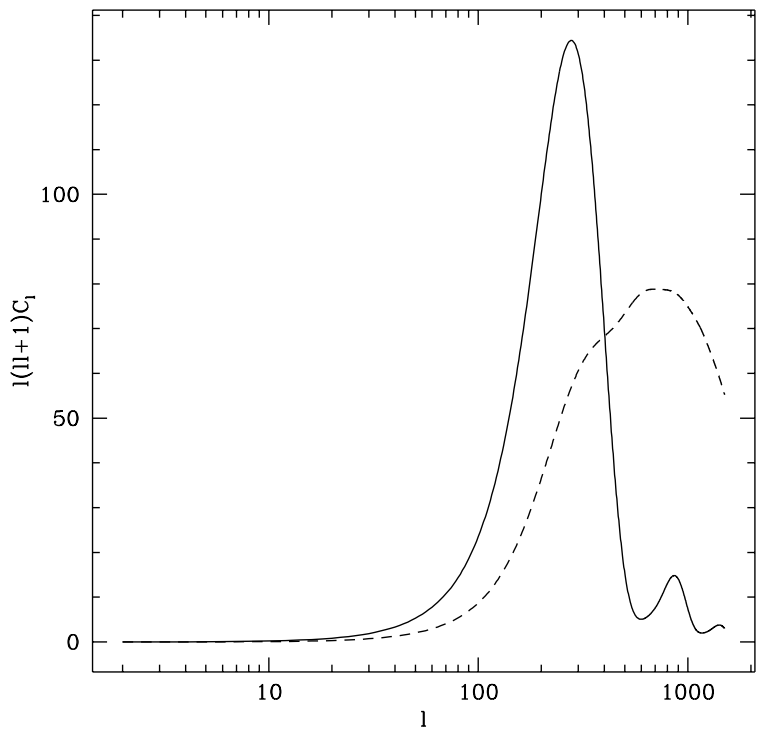}
\caption{
The resulting spectrum of CMB anisotropies from the photon density
perturbations given in Fig.~1 for perfect coherence (solid line)
and complete decoherence (dashed line).}
\end{figure}

\begin{figure}[htb]
\centering
\epsfysize=9.5cm
\epsffile{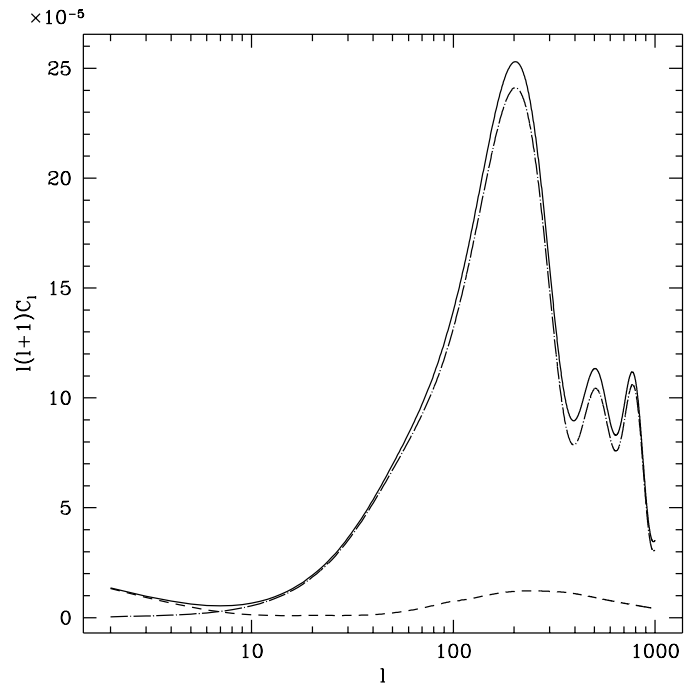}
\epsfysize=9.5cm
\epsffile{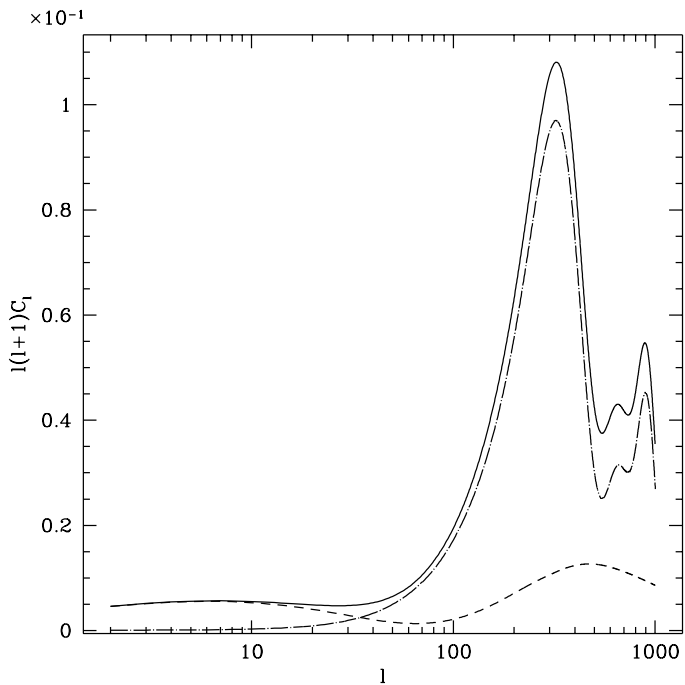}
\caption{
Here and in the subsequent figures, the CMB anisotropies are shown 
in units of $\epsilon^2=(4\pi G\eta^2)^2$. The Sachs Wolfe
contribution alone is indicated by a dashed line and the coherent sum
of acoustic and Doppler terms are shown as dot-dashed curve. The solid
line is the incoherent sum of these two contributions.
Here the seed functions are
determined by the choice $A_1=3$, $A_3=-0.6$ and $A_4=0$. In the top frame
the seed functions decay exponentially, while they decay like a
power law in the bottom one.}
\end{figure}
\begin{figure}[htb]
\centering
\epsfysize=10.5cm
\epsffile{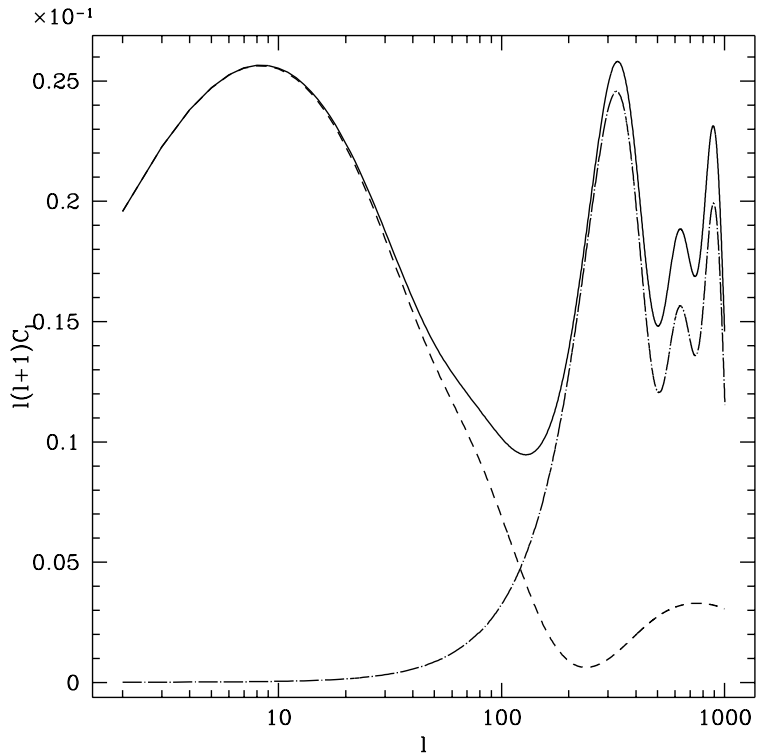}
\caption{
As Fig.~3b, but with $A_3=-0.7$. In this regime ($A_4=0$, $A_3\sim
-(1/4 $ to $ 1/3)A_1$) the resulting spectrum
depends very sensitively on $A_3$. While $A_3=-0.6$ leads to a
perfectly reasonable spectrum with a somewhat high first acoustic
peak, this spectrum is  excluded by observations due to its ``bump'' in
the Sachs Wolfe plateau and the absence of a distinctive acoustic
peak.}
\end{figure}

\begin{figure}[htb]
\epsfysize=7.0cm
\epsffile{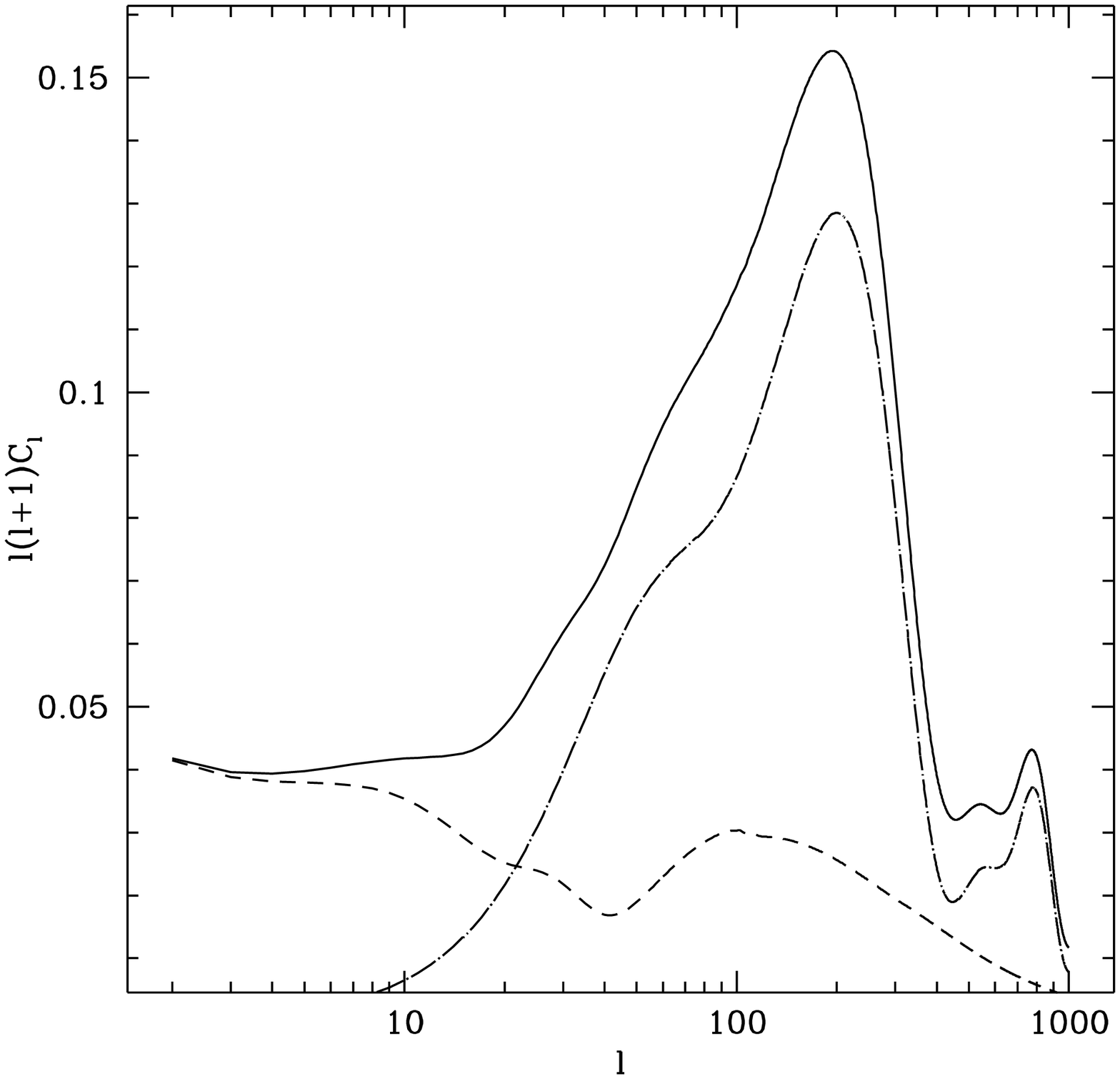}
\epsfysize=7.0cm
\epsffile{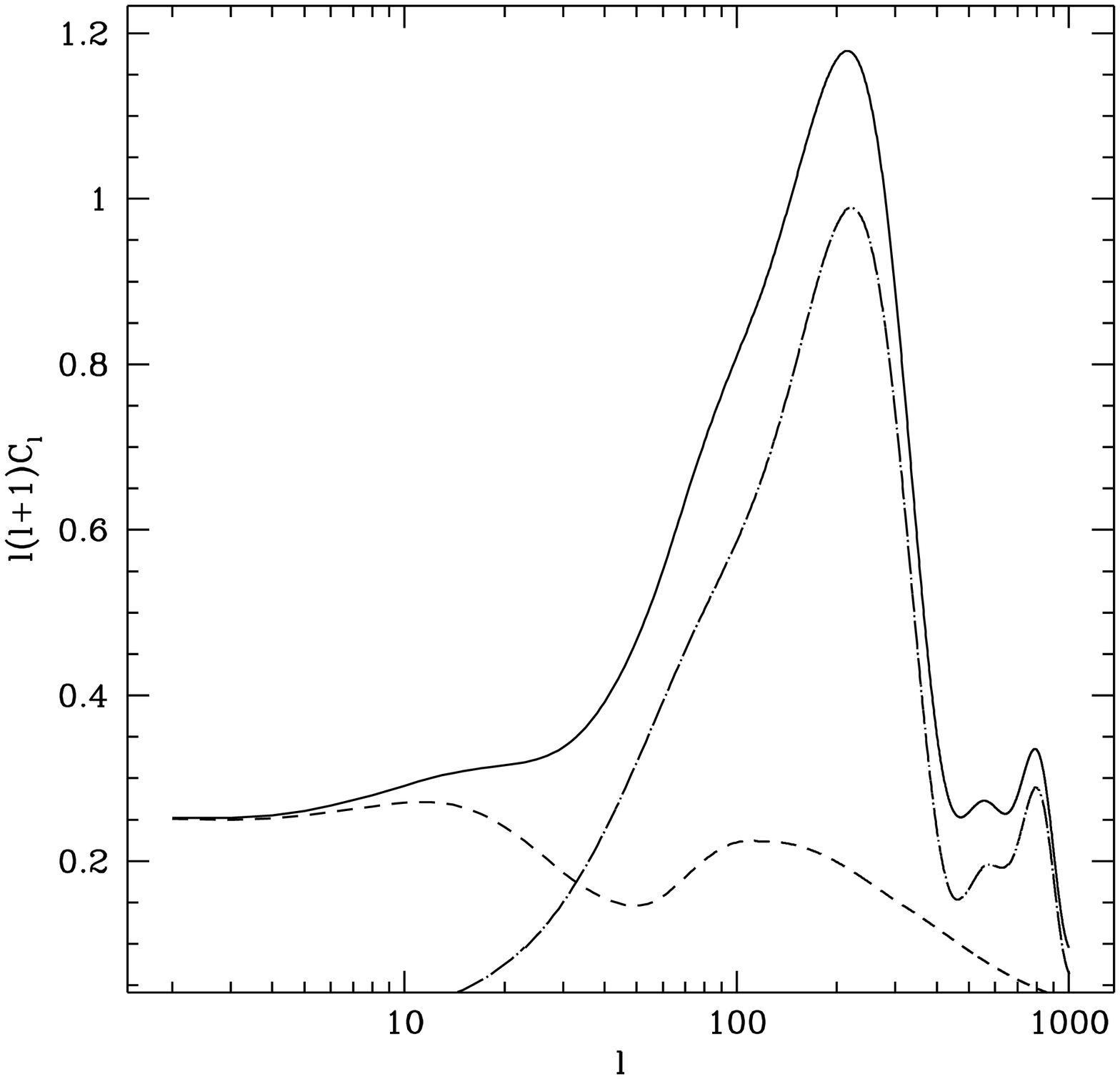}
\centering
\epsfysize=7.0cm
\epsffile{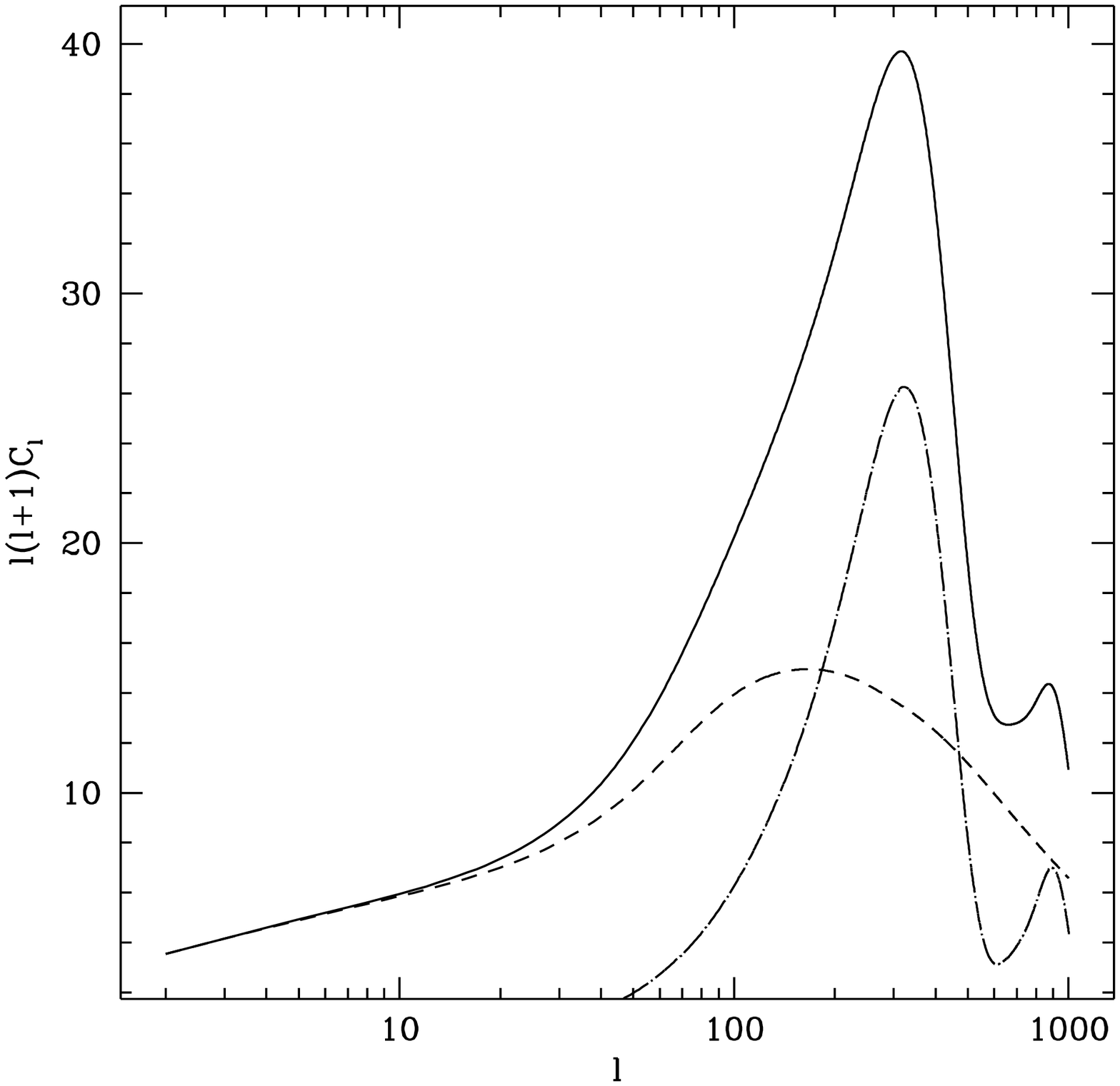} 
\caption{
The CMB anisotropy spectrum for parameters $A_1=3$, $A_3=1$,  $A_4=2$
and exponential decay, $F_i\propto \exp(-x^2)$  (top) and $F_i\propto
\exp(-x^2/2)$  (middle). The corresponding spectrum for seed functions
with power law decay is shown in the bottom frame. The position and relative
amplitude of the first acoustic peak of the spectra (top) and (middle) is
 compatible with an inflationary spectrum. This simple
example hints that it may be possible to ``manufacture'' inflationary
spectra by choosing suitable seed functions.}
\end{figure}

\begin{figure}[htb]
\centering
\epsfysize=10.1cm
\epsffile{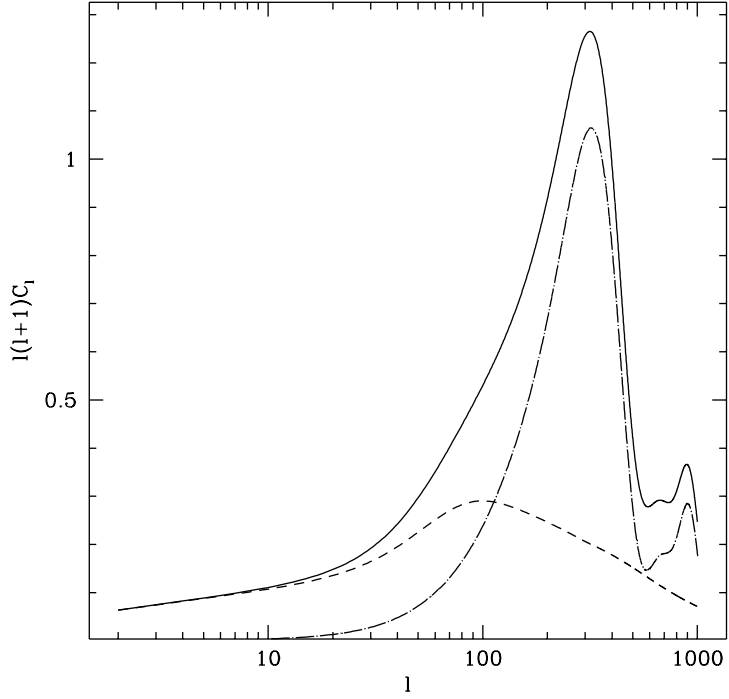}
\epsfysize=10.1cm
\epsffile{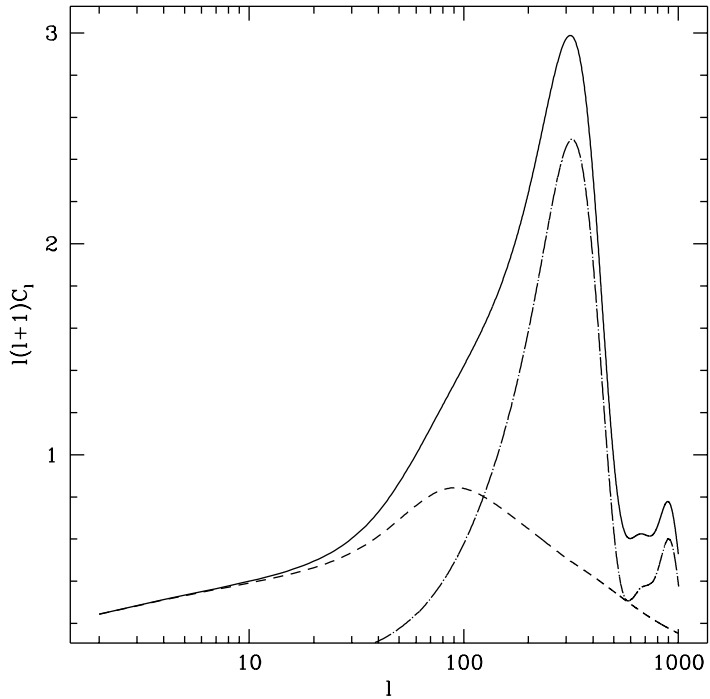}
\caption{
The anisotropy spectra for $A_1=3$, $A_4=A_1/6(2\pi)^2$ 
and $A_3=A_1/18$ (top)
respectively $A_3=-A_1/18$ (bottom) are shown. The seed functions are chosen
to have  power law decay. These are the values $A_i$ which can be
inferred from energy momentum conservation under the assumption of
perfect coherence (see text). The sign of $A_3$ cannot be deduced, but we see
that the results do not depend on it.} 
\end{figure}

\begin{figure}[htb]
\centering
\epsfysize=10.5cm
\epsffile{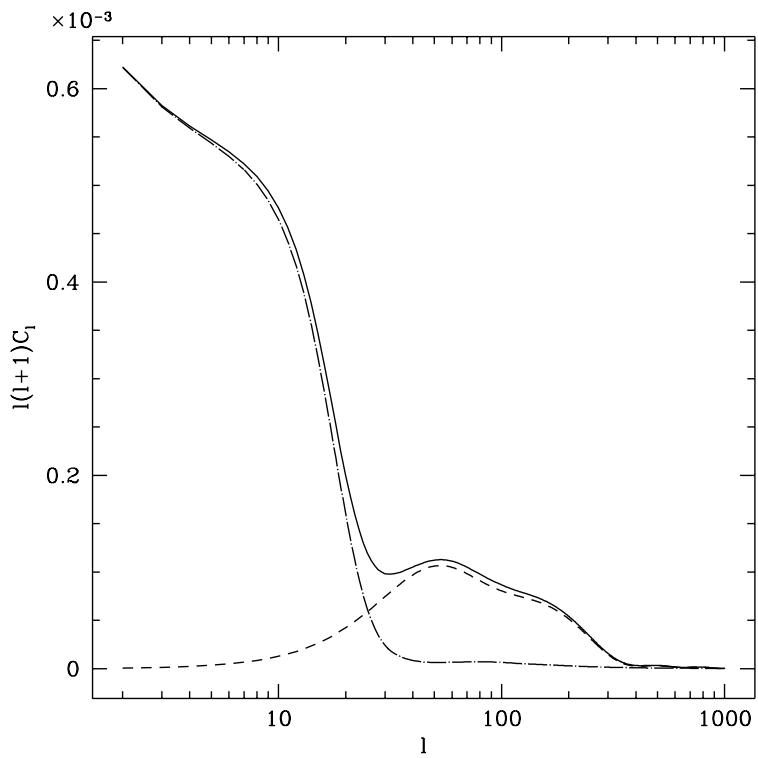}
\caption{
Even extremely strange spectra, like this one with a negative spectral
index and without acoustic peaks can be obtained. For this result
we have chosen exponentially decaying source functions, $F_i\propto
\exp(-x^2)$ and $A_1=3$, $A_3=-1$, $A_4=0$.}
\end{figure}

\begin{figure}[htb]
\centering
\epsfysize=10.5cm
\epsffile{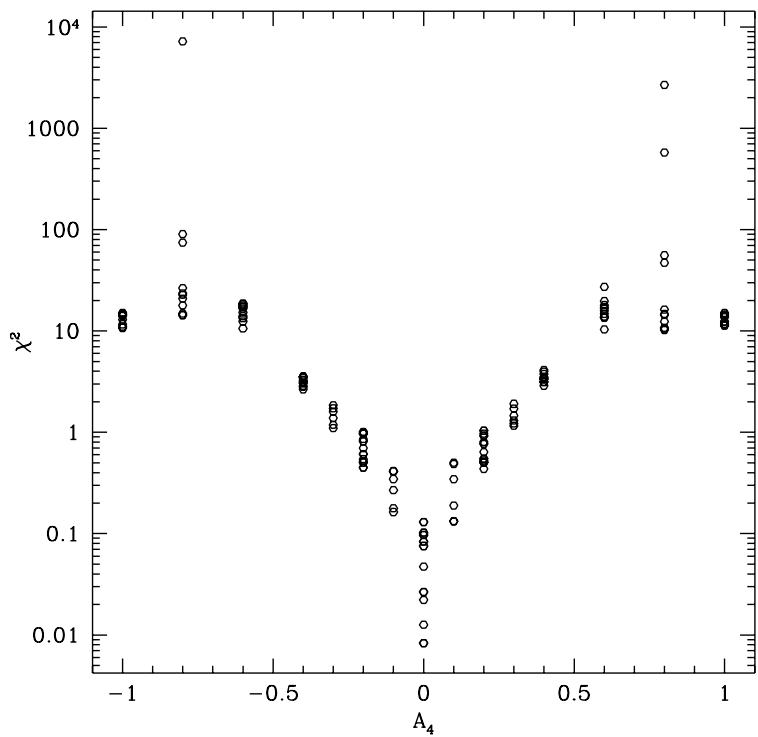}
\caption{
 The CMB anisotropies on large scales, $\ell\le 20$ from 122 
 models with  scaling 
 sources with structure function amplitudes (for details see text) in the 
 regime $-1\le A_4/A_1\le 1$ and $-1\le A_3/A_1\le 1$  are fitted to simple 
 power law spectra with spectral index $-0.5\le n\le 2.5$. The $\chi^2$ 
 of the fit (allowing for 5\% relative error) is shown as function of $A_4$.}
\end{figure}

\begin{figure}[htb]
\centering
\epsfysize=10.1cm
\epsffile{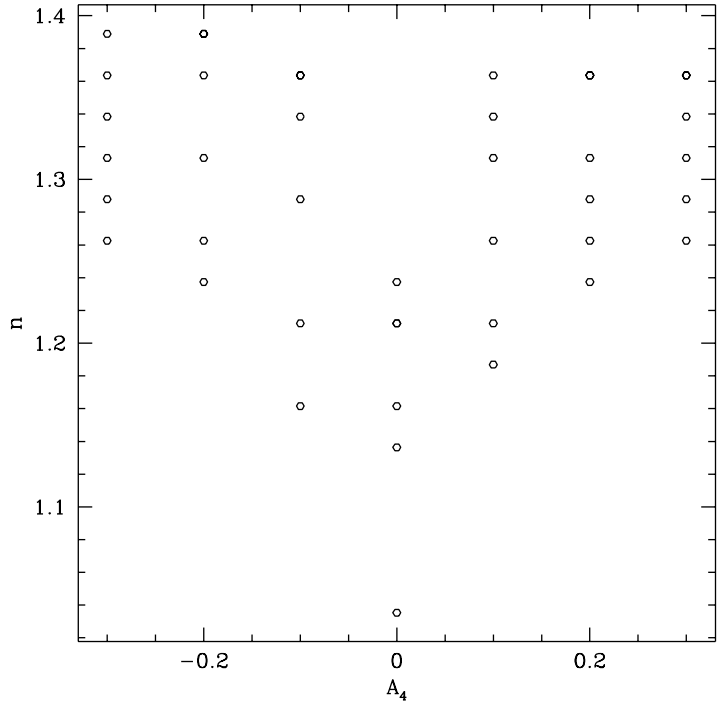}
\epsfysize=10.1cm
\epsffile{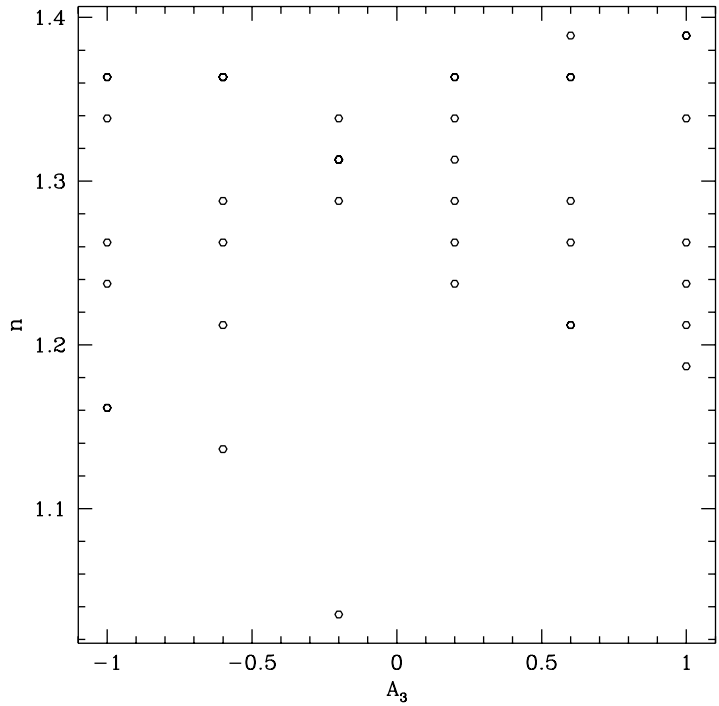}
\caption{
The spectral index for different models with $-0.3\le A_4/A_1\le 0.3$ and
$-1\le A_3/A_1\le 1$ is shown as function of $A_4$ (top). The fact that
there are less than six different  circles visible for some values of
$A_4$ is due to the discrete spacing of about 0.06 in $n$. The bottom
frame shows the same results as a function of $A_3$.}
\end{figure}

\begin{figure}[htb]
\centering
\epsfysize=10.5cm
\epsffile{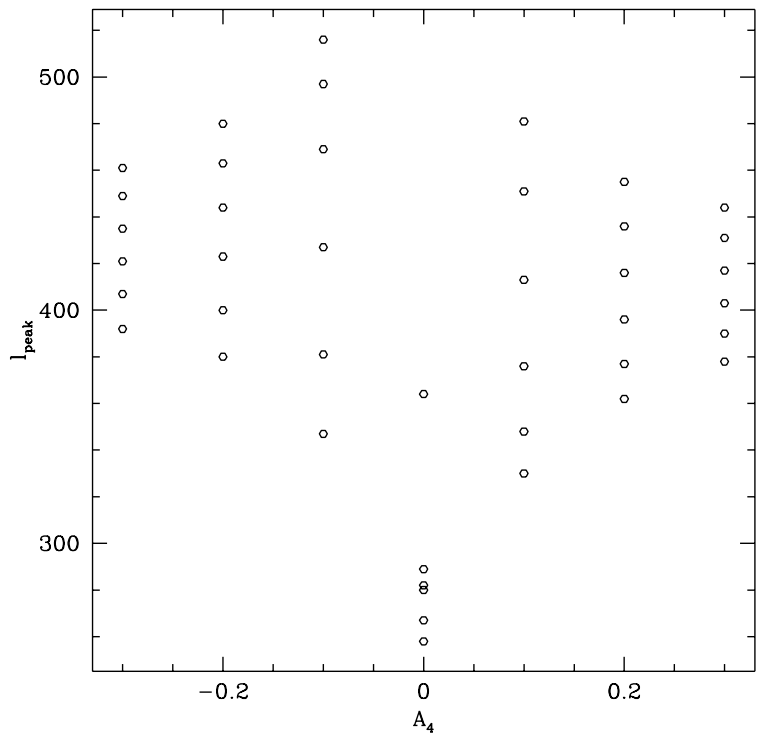}
\caption{
The position of the first acoustic peak
is shown as function of $A_4/A_1$ for different values of $A_3$ in the 
range $-A_1\le A_3\le A_1$.}
\end{figure}

\begin{figure}[htb]
\centering
\epsfysize=10.5cm
\epsffile{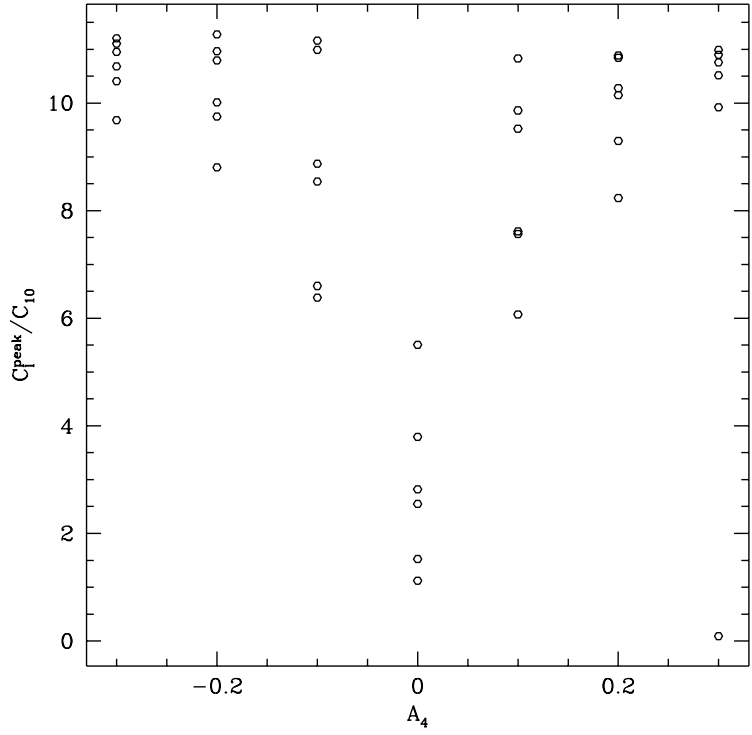}
\epsfysize=10.5cm
\epsffile{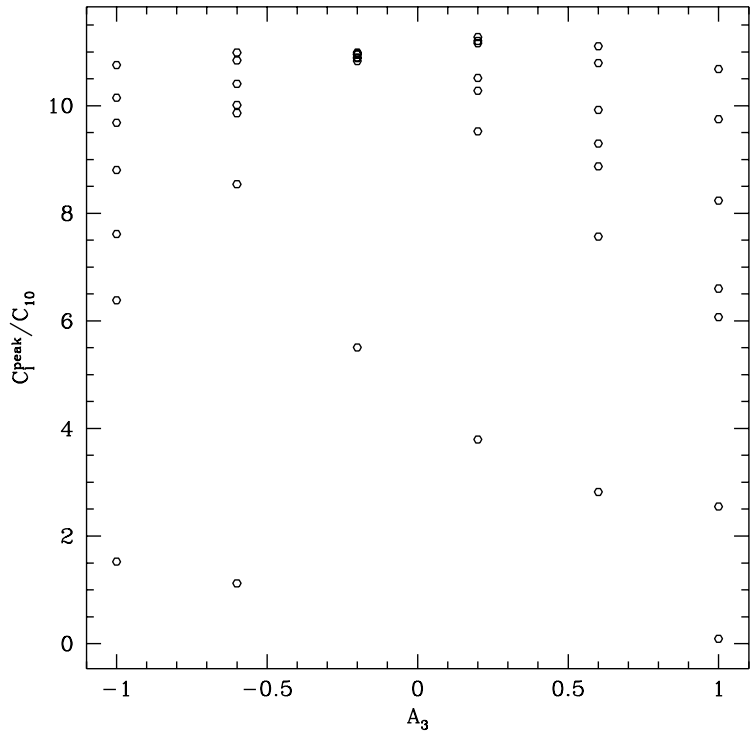}
\caption{
The amplitude of the first acoustic peak over the Sachs Wolfe plateau
is shown as function of $A_4/A_1$ (top) for different values of $A_3$ in the 
range $-A_1\le A_3\le A_1$. The bottom frame shows the same results as
a function of $A_3$.}
\end{figure}

\begin{figure}[htb]
\centering
\epsfysize=10.5cm
\epsffile{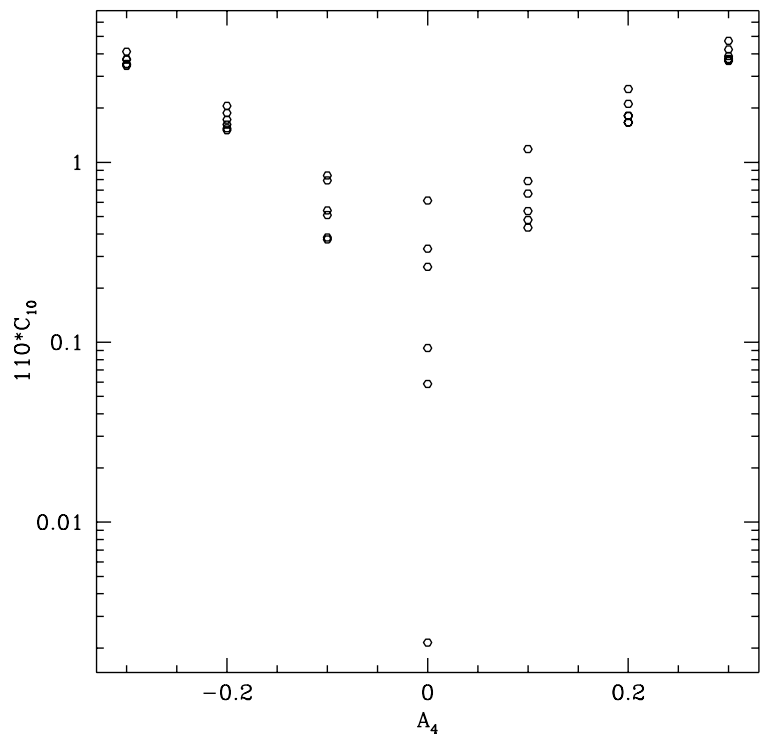}
\caption{
The amplitude $110C_{10}$ is given (in units of $\epsilon^2A_1^2$) as a 
function of the amplitude of anisotropic stresses, $A_4/A_1$. It varies 
over about 3 orders of magnitudes and can become substantially smaller 
than 1, especially for very small values of $A_4$. The significance of 
this finding for the biasing problem is discussed in the text.}
\end{figure}

\end{document}